\documentclass[screen,nonacm,acmtog,review=false]{acmart}

\usepackage{amsmath}
\usepackage{bm}
\usepackage{color,balance,microtype,booktabs}
\usepackage[fleqn,tbtags]{mathtools}
\usepackage{hyperref}
\usepackage[capitalise]{cleveref} 
\usepackage[detect-weight]{siunitx}
\usepackage{pifont,xspace}
\usepackage{multirow}
\usepackage{multicol}
\usepackage{algorithm}
\usepackage{algorithmic}
\usepackage{stfloats}

\graphicspath{{./images/}}

\def \ie {{\emph{i.e}.\thinspace}, }

\newcommand{\name}{\textsc{CGiNS}\xspace}

\newcommand{\revision}{\textcolor[rgb]{0.0,0.0,1.0}}

\AtBeginDocument{%
}

\begin{document}


\title{PCG-Informed Neural Solvers for High-Resolution Homogenization of Periodic Microstructures}

\keywords{Homogenization, Neural Solver, Preconditioned Conjugate Gradient}
\author{Yu Xing}
\affiliation{%
  \institution{Shandong University}
  \city{Qingdao}
  \country{China}
}
\email{xing_yu@mail.sdu.edu.cn}

\author{Yang Liu}
\affiliation{%
  \institution{Microsoft Research Asia}
  \city{Beijing}
  \country{China}
}
\email{yangliu@microsoft.com}

\author{Lipeng Chen}
\affiliation{%
  \institution{Zhejiang Laboratory}
  \city{Hangzhou}
  \country{China}
}
\email{chenlp@zhejianglab.org}

\author{Huiping Tang}
\affiliation{%
  \institution{Hangzhou City University}
  \city{Hangzhou}
  \country{China}
}
\email{thpz@hzcu.edu.cn}

\author{Lin Lu}
\authornote{Corresponding author.}
\affiliation{%
  \institution{Shandong University}
  \city{Qingdao}
  \country{China}
}
\email{llu@sdu.edu.cn}

\begin{abstract}
  The mechanical properties of periodic microstructures are pivotal in various engineering applications. Homogenization theory is a powerful tool for predicting these properties by averaging the behavior of complex microstructures over a representative volume element. However, traditional numerical solvers for homogenization problems can be computationally expensive, especially for high-resolution and complicated topology and geometry. Existing learning-based methods, while promising, often struggle with accuracy and generalization in such scenarios.

To address these challenges, we present \name, a \revision{geometry-aware} preconditioned conjugate gradient solver informed neural network for solving homogenization problems. \name leverages sparse and periodic 3D convolution to enable high-resolution learning while ensuring structural periodicity. It features a multi-level network architecture that facilitates effective learning across different scales and employs minimum potential energy as label-free loss functions for self-supervised learning. The integrated preconditioned conjugate gradient iterations ensure that the network provides PCG-friendly initial solutions for fast convergence and high accuracy. 
Additionally, \name imposes a global displacement constraint to ensure physical consistency, addressing a key limitation in prior methods that rely on Dirichlet anchors.

Evaluated on large-scale datasets with diverse topologies and material configurations, \name achieves state-of-the-art accuracy (relative error below $1\%$) and outperforms both learning-based baselines and GPU-accelerated numerical solvers. Notably, it delivers $10\times$ speedups over traditional methods while maintaining physically reliable predictions at $512^3$ resolution.

\end{abstract}

\maketitle

\section{Introduction}
\label{sec:intro}

Microstructures, prevalent in natural materials, have garnered significant interdisciplinary attention for their ability to achieve superior mechanical performance at minimal mass density, particularly in applications requiring high strength-to-weight and stiffness-to-weight ratios \cite{brandon2013microstructural}. 
\revision{Periodic single-phase lattice microstructures constitute the dominant class implemented in additive-manufacturing (AM)–driven design and production: using a single certified feedstock simplifies fabrication and qualification, yields repeatable, designable effective properties, and scales reliably to large parts and databases of candidates \cite{askari2020additive,bajaj2020steels,liu2022additive,tan2020microstructure}.
As a result, single-phase periodic cells are the default choice in lightweighting, energy-absorption, and biomedical implant design~\cite{li1996microstructure}.}

To meet the demands of these applications, researchers have developed various modeling methodologies tailored to specific requirements. For instance, triply periodic minimal surface (TPMS)-based lattice structures offer biomimetic advantages, such as smooth curvature gradients and bi-continuous solid-void topologies, making them ideal for thermal exchangers and osseointegrated biomedical implants \cite{al2020functionally,xu2023new}. Shell lattices with curvature-driven architectures enable multifunctional performance in thermal radiation and permeability-critical systems \cite{liu2022parametric}, while plate lattice microstructures exhibit stiffness metrics that approach theoretical bounds, making them suitable for energy-absorbing systems \cite{sun2023parametric}. Truss-based microstructures, on the other hand, can be engineered to exhibit auxetic behaviors, such as negative Poisson's ratios, through careful nodal connectivity design \cite{panetta2015elastic,zheng2023unifying}.

Across these design families, accurate and efficient evaluation of effective mechanical properties remains essential for guiding structure selection and downstream optimization~\cite{liu2021mechanical}, motivating the development of computational tools capable of resolving mechanical responses across increasingly complex, high-resolution microarchitectures.

Homogenization theory provides a foundational framework for linking microscale material behavior to macroscopic effective properties in periodic media. At its core, homogenization reduces to solving an elasticity boundary value problem -- \ie a system of partial differential equations (PDEs) that enforce mechanical equilibrium under periodic boundary conditions~\cite{babuvska1976homogenization,terada2000simulation,bensoussan2011asymptotic,li1996microstructure,torquato2002random}. By resolving these PDEs on representative unit cells, homogenization enables the derivation of continuum-scale constitutive laws without explicitly modeling the full microstructure. 
\revision{In practice, multiple computational strategies have been developed. Fast Fourier transform (FFT)-based approaches are efficient for regular-grid periodic problems~\cite{schneider2021review,chen2020comparison}, but they suffer from convergence issues and numerical oscillations when applied to high-contrast microstructures~\cite{schneider2016computational,lucarini2022adaptation}.
Theoretical homogenization methods, such as asymptotic expansion or mean-field theories~\cite{sekkate2022elastoplastic}, provide analytical approximations rather than fully numerical solutions.}
Among available techniques, the finite element method (FEM) remains the most widely adopted approach due to its mesh flexibility and robustness across complex topologies~\cite{omori2007overview,keyes2013multiphysics,solin2006partial}. However, FEM-based homogenization typically involves the solution of large-scale sparse linear systems, where fine discretization required for accuracy imposes considerable computational and memory costs~\cite{barkanov2001introduction,saeb2016aspects}. 
Iterative solvers, such as conjugate gradient (CG) methods and multigrid solvers, are often employed to alleviate computational costs through GPU acceleration, but they remain slow to converge, particularly for high-resolution discretizations.

\begin{figure*}[b]
    \centering
    \includegraphics[width=\linewidth]{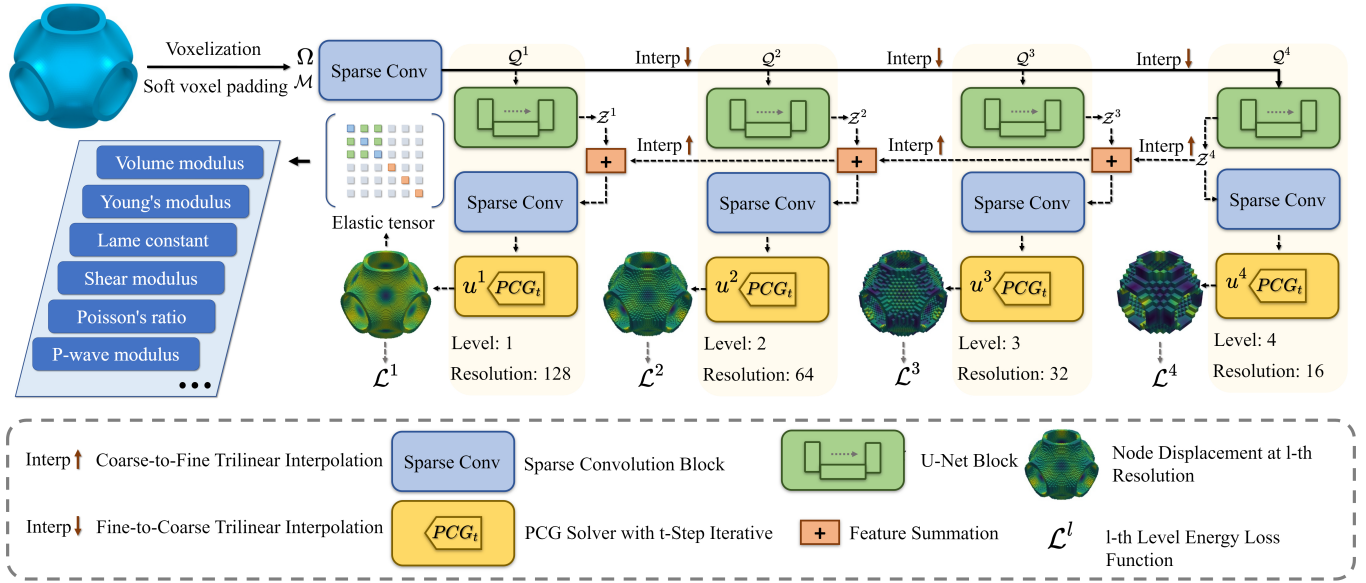}
    \caption{\revision{\textbf{Overview of the \name framework.}This framework is illustrated using periodic lattice structures with a target resolution of $n$. Given a periodic microstructure, \name voxelizes the input domain and applies soft-voxel padding to enhance boundary fidelity. The geometry and base material tensor are processed through a sparse convolutional encoder to extract multi-resolution features $\mathcal{Q}^l$. At each level $l$, a U-Net block refines these features, which are then decoded into displacement predictions $u^l$ and corrected via global translation to satisfy displacement constraints. A $t$-step preconditioned conjugate gradient (PCG) smoother eliminates high-frequency errors. Coarse-to-fine trilinear interpolation propagates feature information across levels, while energy-based losses $\mathcal{L}^l$ are computed at each resolution for self-supervised training. All computations operate on sparse tensors to ensure scalability to high-resolution structures.}}
    \label{fig:Pipeline}
\end{figure*}

Deep learning has demonstrated significant potential for solving PDEs, offering faster inference and greater flexibility than traditional numerical methods~\cite{beck2020overview,beck2021solving}. However, the field remains in its early stages, with several unresolved challenges. Current research can be broadly categorized into two major directions.

\paragraph{End-to-End Neural Solutions} These approaches aim to fully replace traditional solvers by directly approximating PDE solutions using techniques such as physics-informed neural networks (PINNs)~\cite{raissi2019physics,cai2021physics,cai2021physicsheat}, neural operators (Fourier neural operator and DeepONet)~\cite{li2020fourier,lu2021learning,lu2019deeponet,zappala2024learning,cao2024laplace}, and emerging attempts to leverage large language models (LLMs) for PDE analysis~\cite{zhou2024unisolver,bhatnagar2025equations}. While promising, these methods are primarily tested on relatively simple PDEs such as Poisson equations. 
\revision{Recent works have explored learning-based homogenization at the operator level~\cite{bhattacharya2023learning, bhattacharya2024learning}, focusing on constructing reduced models of homogenized PDEs in idealized settings. In contrast, our work targets a solver-level challenge with industrial-grade accuracy: directly solving high-resolution 3D homogenization PDEs on voxelized microstructures under periodic boundary conditions, while embedding physical constraints and iterative solver consistency.}
Furthermore, neural operator frameworks often require large-scale labeled datasets for training, and generating such data -- especially for high-resolution or physics-specific problems -- can be computationally prohibitive.

\paragraph{Hybrid Approaches} These methods integrate neural networks with numerical solvers to enhance performance while retaining the robustness of traditional methods. Neural networks have been applied to accelerate intermediate solution steps~\cite{luz2020learning,chen2022meta,han2024ugrid}, predict efficient preconditioner~\cite{trifonov2024learning,azulay2022multigrid,zhang2024blending}, and improve initial guesses for iterative solvers~\cite{um2020solver,zou2025learning,huang2020int,pestourie2023physics}. 
\revision{However, most of these approaches essentially operate from a matrix perspective, which inevitably discards  underlying geometric information. We argue that the nature of this problem is a combination of geometry and numerics, and an effective solver should couple the two aspects more tightly. In our approach, the neural network component captures geometric characteristics and enforces problem-specific constraints such as periodicity, while the numerical part adopts PCG and iterative schemes. In this way, the network provides high-quality initial guesses to the numerical solver, and joint training further strengthens their coupling.}


Existing neural network-based methods for microstructure homogenization, such as PH-Net~\cite{peng2022ph} and Label-free~\cite{zhu2024learning}, directly approximate solutions to linear systems from geometric inputs. While effective on simple lattice structures, these models degrade significantly on complex topologies, often exhibiting instability and reduced accuracy (as shown in \cref{sec:2.2}). Most results are reported on small datasets without large-scale evaluation, limiting generalization. Current models also struggle with high-resolution inputs due to memory and compute bottlenecks. Boundary conditions are typically imposed through weak supervision, such as loss penalties, rather than being explicitly encoded into the model architecture. In particular, Dirichlet conditions are often enforced by arbitrarily fixing a single node (anchor node) to zero displacement -- an approach that may be sufficient for regular geometries but becomes unreliable for topologically complex microstructures, where the anchor point cannot be consistently or meaningfully inferred. This often leads to violation of physical constraints and further amplifies prediction error. We also found that the 
reported efficiency gains are often based on biased hardware comparisons and convergence conditions, similar to the observations by \cite{mcgreivy2024weak}. 
These deficiencies highlight the core challenges in neural microstructure homogenization: achieving robust high-accuracy predictions across diverse and complex structures, ensuring boundary condition consistency, scaling to high resolutions, and maintaining computational efficiency without sacrificing generalization.

In this work, we present a novel conjugate-gradient-solver-informed neural framework -- \name, designed for the rapid and scalable simulation of homogenized linear elastic properties in periodic microstructures. \cref{fig:Pipeline} illustrates the overall architecture of \name. Leveraging sparse 3D convolutions and periodic encoding, \name predicts numerical solutions to homogenization PDEs directly from voxelized geometries and base material parameters, scaling up to resolutions of $512^3$.

To ensure convergence and physical consistency, \name embeds preconditioned conjugate gradient (PCG) iterations into the network architecture, enabling high-quality initializations that accelerate convergence while maintaining solver stability. \revision{We use projection energy to quantify the alignment between the network initialization and the preconditioned CG system, providing key evidence for rapid convergence with only a few iterations.} The hybrid multi-level framework supports hierarchical feature extraction and resolution-aware interpolation, improving accuracy across structural scales. Additionally, a physics-informed variational loss based on minimum potential energy enables self-supervised training without labeled displacement fields. A global displacement constraint is also introduced to enforce boundary correctness without relying on heuristic anchors.

To enable reproducible and rigorous evaluation, we construct three benchmark datasets featuring diverse geometric topologies and mechanical behaviors. Across all datasets, \name achieves relative elasticity tensor errors below $1\%$, while outperforming both learning-based and GPU-accelerated numerical solvers by $2\times$–$10\times$ in runtime -- demonstrating its superior accuracy, generalization, and computational efficiency.

\section{Results}
\label{sec:result}

\subsection{Overview of \name}

\paragraph{Problem formulation}
Let $\Omega \subset \mathbb{R}^3$ denote a representative volume element (RVE) of a periodic microstructure with prescribed geometry and boundary conditions, and let $\mathcal{M}$ represent the underlying base material properties. Our objective is to compute the effective linear elastic response of the heterogeneous medium, expressed as the homogenized elasticity tensor $C^H \in \mathbb{R}^{6 \times 6}$.

According to classical homogenization theory~\cite{andreassen2014determine,dong2019149}, the macroscopic constitutive behavior of the material can be described as:
\begin{equation}
C^H = \mathcal{H}_{\text{omo}}(\Omega, \mathcal{M}),
\label{eq:2-1}
\end{equation}
where $\mathcal{H}_{\text{omo}}(\cdot)$ is a deterministic mapping defined by the solution of a boundary value problem derived from the equations of linear elasticity.

In practice, computing $C^H$ requires solving the microscopic equilibrium equations governed by the linear elasticity PDE under periodic boundary conditions. Discretizing the PDE using the finite element method (FEM) leads to the linear system:
\begin{equation}
K(\Omega, \mathcal{M}) u = f(\Omega, \mathcal{M}),
\label{eq:2-2}
\end{equation}
where $K \in \mathbb{R}^{3n \times 3n}$ is the global stiffness matrix, $f \in \mathbb{R}^{3n}$ is the external load vector, and $u \in \mathbb{R}^{3n}$ is the node displacement vector. The homogenized tensor $C^H$ is then obtained by post-processing the computed displacement field $u$.

Solving \cref{eq:2-2} is computationally expensive, particularly for high-resolution or topologically complex microstructures. To address this challenge, we propose \name. (Further details on the homogenization setup and symbol definitions are provided in Supplementary.)

\paragraph{Overview of the \name framework}
\cref{fig:Pipeline} illustrates the architecture of \name. The input microstructure is voxelized with soft voxel padding to enhance surface representation. Geometric and material properties are encoded via sparse convolutions into hierarchical features. A U-Net backbone transforms these features across resolutions using trilinear interpolation. At each level, displacement predictions are decoded, normalized via a global translation constraint, and refined via a $t$-step preconditioned conjugate gradient iterations (PCG module). Importantly, no ground-truth labels are required--\name is trained using physically grounded energy-based loss at each resolution, ensuring stable convergence even for complex topologies. All modules are implemented in sparse format, enabling efficient training on resolutions up to $512^3$.

More details about \name and the loss function can be found in \cref{sec:method} and Supplementary.


\paragraph{Training and benchmark datasets} 
We conduct all training and evaluation using three representative microstructure datasets (TPMS, PSL, and Truss) with increasing geometric complexity. The TPMS dataset contains triply periodic minimal surface structures characterized by smooth, continuous morphologies and is widely used due to its regularity and predictable mechanics. In contrast, the PSL and Truss datasets exhibit complex topologies with sharp geometric variations and connectivity discontinuities, providing more realistic and challenging scenarios.

Each dataset is constructed with diverse volume fractions ranging from $10\%$ to $40\%$, and elastic properties sampled from physically plausible distributions. The data are split into disjoint training and test sets, with 30,000–78,000 samples for training and 3,000 samples for testing per dataset, as detailed in Supplementary. All reported results are evaluated on the held-out test sets, ensuring that benchmarking reflects generalization to unseen microstructures and material configurations.

\subsection{Accuracy and robustness of \name}
\label{accuracy}

\paragraph{Performance across structural complexity} 
\label{sec:2.2}
To evaluate the accuracy and robustness of \name under increasing geometric complexity, we benchmark it against three representative learning-based baselines: 3D-CNN~\cite{rao2020three}, Label-Free~\cite{zhu2024learning}, and PH-Net~\cite{peng2022ph}. Among these, 3D-CNN directly regresses the effective elasticity tensor $C^H$ from geometry, whereas Label-Free, PH-Net, and \name follow a two-stage strategy by first predicting the displacement field $u$ and then deriving $C^H$ via \cref{eq:2-1}.

To ensure consistent evaluation, all models were tested on the same datasets with identical resolution ($n=64$), and were assessed using two complementary metrics:
\begin{align}
        &\delta = \frac{\Vert C^H-\widetilde{C}^H\Vert_F}{\Vert \widetilde{C}^H\Vert_F},\\
        &\Vert r\Vert  = \Vert f-Ku\Vert_2,
\label{eq:2-5}
\end{align}
where $\delta$ is the relative error of the predicted elasticity tensor, and $\Vert r \Vert$ is the residual of the linear system. The reference solution $\widetilde{C^H}$ was obtained using the AmgX solver with $\Vert r \Vert < 10^{-5}$.
\begin{table}[t]
        \caption{ \textbf{Performance of \name on different structures.} We compare \name with SOTA learning-based approaches. To accommodate the resolution limit of all methods, we set $n=64$. The volume fraction of the test models ranges from $10\%$ to $40\%$, with each subset containing 3000 microstructure models. Since related works do not universally support different material configurations, both the test and training sets are configured with the same homogeneous material. $\delta$ denotes the relative error (in percentage, $\%$). \textbf{$\Vert r \Vert$denotes the residual, divided by $10^{-3}$}. For each test task, we report the maximum error $\delta_{Max}$, average error $\delta_{Mean}$, maximum residual $\Vert r \Vert_{Max}$, and average residual $\Vert r \Vert_{Mean}$.}
    \scalebox{0.9}{
    \begin{tabular}{llrrrr}
        \toprule
        \textbf{Dataset} &\textbf{ Network} & $\bm{\delta_{Max} }$  &$\bm{\delta_{Mean}}$ & $\bm{\Vert r\Vert_{Max}}$ &$\bm{\Vert r\Vert_{Mean}}$\\
         \midrule
         \multirow{4}{*}{TPMS} &
          3D-CNN~\cite{rao2020three}  &$17.57\%$  &$8.54\%$ &/&/ \\
         & Label-Free~\cite{zhu2024learning} &$44.81\%$  & $23.51$&$26.63$&$21.12$\\
         & PH-Net~\cite{peng2022ph}  &$2.62\%$  &$1.45\%$ &$16.22$ &$12.47$\\
         & \name  &$\bm{0.18\%}$  & $\bm{0.05\%}$ &$\bm{1.48}$ &\textbf{1.16}\\
       \midrule
         \multirow{4}{*}{PSL} &
          3D-CNN~\cite{rao2020three} & $111.99\%$ & $40.16\%$&/&/ \\
       &   Label-Free~\cite{zhu2024learning} & $239.98\%$&$87.51\%$&$39.78$ &$32.47$ \\
       &PH-Net~\cite{peng2022ph} & $47.34\%$ &$ 23.19\%$ &$25.90$ &$17.36$\\
       &   \name & $\bm{0.64\%}$ & $\bm{0.31\%}$ & $\bm{3.04}$ & $\bm{2.06}$\\
      
         \midrule
         \multirow{4}{*}{Truss} &
          3D-CNN~\cite{rao2020three} &$197.54\%$ & $47.13\%$ &/&/\\
       & Label-Free~\cite{zhu2024learning} &$836.25\%$&$177.47\%$ &$60.99$ &$48.64$ \\
       &PH-Net~\cite{peng2022ph} & $357.14\%$&$65.67\%$&$25.20$ &$15.62$\\
       &    \name & $\bm{0.84\%}$& $\bm{0.51\%} $& \textbf{8.04 }&\textbf{3.46}\\
       \bottomrule
    \end{tabular}
    }
    \label{tab:structure}
\end{table}
As shown in \cref{tab:structure}, \name consistently delivers the highest accuracy across all datasets. On TPMS structures, where displacement fields are smooth, all models achieve stable performance. \name still outperforms alternatives with $\delta_{Mean} = 0.05\%$ and lowest residuals.
On structurally complex datasets such as PSL and Truss, competing models show significant degradation. 3D-CNN fails to generalize, yielding mean errors over $40\%$ and frequent instability. Label-Free and PH-Net reduce residuals to $\mathcal{O}(10^{-2})$, but their predictions remain noisy and inconsistent, with $\delta > 10\%$ in most cases. In contrast, \name achieves robust predictions with $\Vert r \Vert = \mathcal{O}(10^{-3})$ and $\delta < 1\%$ across all samples.

Furthermore, \name supports high-resolution inference up to $n=512$, while competing methods are fundamentally limited to low-resolution settings due to memory constraints or reliance on dense tensor operations.

\paragraph{Robustness across material configurations}
To evaluate whether \name can generalize across varying base material parameters, we assess its predictive stability under randomized isotropic elasticity conditions.
Unlike prior learning-based methods that operate under fixed material assumptions, \name is designed to support a continuous spectrum of isotropic base materials. By explicitly encoding the elasticity tensor $C^b$ as input, the model adapts its predictions based on both geometry and material composition, enabling accurate and physically consistent inference across varying configurations.

To evaluate this capability, we test \name on three datasets -- TPMS, PSL, and Truss--where each structure is paired with randomly sampled material properties (details provided in Supplementary.)

As shown in \cref{tab:material}, \name maintains high accuracy and numerical stability across all materials and structures, with mean relative tensor error below $0.5\%$ and residual norms consistently at $\mathcal{O}(10^{-3})$.

\subsection{Runtime efficiency compared to linear equation solver}
\paragraph{GPU-based numerical solver}
To evaluate the computational efficiency of \name, we compare it against two GPU-accelerated multigrid solvers: NVIDIA AmgX~\cite{AMGX}, which implements algebraic multigrid (AMG), and Homo3D~\cite{zhang2023}, which incorporates geometric multigrid (GMG) strategies with CUDA optimization. These solvers decompose computation into two phases: system matrix assembly (Init.) and iterative solution via preconditioned conjugate gradient (Cal.).

For fair benchmarking, all methods are evaluated on the same datasets under identical discretization and solver settings. The convergence threshold for each numerical method is standardized by matching the residual norm $\Vert r \Vert$ achieved by \name, ensuring consistency in solution accuracy. Degrees of freedom (DOF) are reported based on the number of unknowns in the discretized system.

\begin{table}[t]
    \centering
    \caption{ \textbf{Performance of \name on different structures and different materials. }
    The volume fraction of the test models ranges from $10\%$ to $40\%$, with each subset containing 3000 microstructure models. \textbf{$\Vert r \Vert$denotes the residual, divided by $10^{3}$}. For each group of test data, we record the maximum error $\delta_{Max}$, average error $\delta_{Mean}$, maximum residual $\Vert r \Vert_{Max}$, and average residual $\Vert r \Vert_{Mean}$.}
        \begin{tabular}{lrrrrr}
        \toprule
        \textbf{Dataset }   & $\bm{\delta_{Max}}$ &$\bm{\delta_{Mean}}$  &$\bm{\Vert r\Vert_{Max}}$ & $\bm{\Vert r\Vert_{Mean}}$  \\
        \midrule
        TPMS   &$0.16\%$  & $0.04\%$ &$1.32$&1.02\\
        PSL  & $0.72\% $ &$ 0.28\%$ & $5.54$ &$2.14$\\
        Truss & $0.91\%$  & $0.45\%$ & $9.94$ & $3.38$\\
       \bottomrule
    \end{tabular}
    
    \label{tab:material}
 \end{table}

\begin{table*}
    \centering
   \caption{\revision{\textbf{Runtime and resolution scaling across solvers.} We compare \name\ with two \textit{purely numerical GPU multigrid} solvers—AmgX (AMG) and Homo3D (GMG), and a \textit{numerical–neural hybrid} method, GNN-AMG~\cite{luz2020learning}. 
All methods are evaluated on identical datasets with matched degrees of freedom (DOF), and the residual convergence thresholds are aligned to those reached by \name\ for fair comparison.
"Init." denotes the time required to assemble system matrices for AmgX, stencil preparation for Homo3D, graph feature building for GNN-AMG and feature encoding for \name. 
Cal. refers to solver execution time: for AmgX and Homo3D, the time of their iterative solves; for GNN-AMG, the time of the AMG iterations; and for \name, the time of neural inference plus PCG iterations.
For GNN-AMG, \textbf{S-rate} is the \emph{success rate}, i.e., the fraction of test instances that converge to the prescribed tolerance within the iteration budget.}}
    \scalebox{.85}{
    \begin{tabular}{ccrrrr|rrr|rrrr|rrrl}
        \toprule
         \multirow{2}{*} {\textbf{Dataset} } & \multirow{2}{*} {\textbf{Res.} }  & \multirow{2}{*}{\textbf{DOF}} & \multicolumn{3}{c}{\textbf{AmgX}~\cite{AMGX}} & \multicolumn{3}{c}{\textbf{Homo3D}~\cite{zhang2023}} &\multicolumn{4}{c}{\textbf{GNN-AMG}~\cite{luz2020learning}} & \multicolumn{4}{c}{\textbf{\name}}\\
         & & & \multicolumn{1}{c}{\textbf{Init. (s)}} & \multicolumn{1}{c}{\textbf{Cal.}}&\multicolumn{1}{c}{\textbf{Total}} & \multicolumn{1}{c}{\textbf{Init.}} & \multicolumn{1}{c}{\textbf{ Cal.}}&\multicolumn{1}{c}{\textbf{Total}}& \multicolumn{1}{c}{\textbf{Init.}} & \multicolumn{1}{c}{\textbf{ Cal.}}&\multicolumn{1}{c}{\textbf{Total}} &\multicolumn{1}{c}{\textbf{S-rate}}&  \multicolumn{1}{c}{\textbf{Init.}} & \multicolumn{1}{c}{ \textbf{Cal.}}&\multicolumn{1}{c}{\textbf{Total}}&\multicolumn{1}{c}{\textbf{Improve.}}\\

        \midrule
        \multirow{4}{*}{TPMS} &\multirow{1}{*}{64}  & $251K$ &0.04 &0.27 &0.32 &0.62 &0.25& 0.87 &0.05&0.23&0.28& $87.11\%$ & 0.05 & 0.09 &\textbf{0.14 }&$2.28\times$ / $6.21\times$ / $2.00\times$\\
        & \multirow{1}{*}{128} & $2M$ &0.30 &3.10 &3.40 &0.67 & 1.04&1.71 & 0.28 & 2.90 & 3.18 & $80.77\%$ &0.28 & 0.55 & \textbf{0.82} &$4.11\times$ / $2.06\times$ / $3.87\times$\\
        & \multirow{1}{*}{256}  & $16M$ &1.27& 7.84& 9.10&0.76 & 6.71 &7.46 &/&/&/&/& 0.02 &3.30&\textbf{3.33} &$2.73\times$ / $2.24\times$ \\
        & \multirow{1}{*}{512} &$128M$ & 2.28 & 80.52 & 82.80 &1.04 & 66.92& 67.96&/&/&/&/&0.28 &7.58 &\textbf{7.86} & $10.24\times$ / $8.64\times$\\
        
        \midrule
        \multirow{4}{*}{PSL} 
        &\multirow{1}{*}{64}  & $275K$ &0.05 &0.44 &0.49 &0.62 &0.37 &0.99 &0.06&1.07&1.13& $20.02\%$ & 0.06 & 0.09 & \textbf{0.15} & $3.11\times$ / $6.23\times$ / $7.53\times$\\
       
        & \multirow{1}{*}{128} & $2M$ &0.36 &3.10&3.46 &0.67 & 1.16&1.83 &0.30&6.04 & 6.34 & $10.48\%$ & 0.30 & 0.57 & \textbf{0.87 }& $3.97\times$ / $2.10\times$ / $7.28\times$\\
        
        & \multirow{1}{*}{256}  & $17M$ &1.32 & 15.61& 16.93 &0.75 & 6.28& 7.03 &/&/&/&/&0.02 &  3.30 & \textbf{3.32} &$5.09\times$ / $2.12\times$\\
        
        & \multirow{1}{*}{512}  & $136M$  &2.34 & 120.50 & 122.84 & 1.04& 80.25 &81.29 &/&/&/&/&0.34 &10.88 &\textbf{11.22} &$10.77\times$ / $7.24\times$\\
       \midrule
       
        \multirow{4}{*}{Truss} &\multirow{1}{*}{64} & $253K$ & 0.05 &0.41 & 0.46 &  0.62& 0.39 & 1.01 &0.06&1.30&1.36&$23.55\%$& 0.06&0.09 & \textbf{0.15}&$2.95\times$ / $6.38\times$ / $9.07\times$ \\
       
        & \multirow{1}{*}{128}  & $2M$ &0.30 &3.22 & 3.52 &0.67 &1.17 & 1.84 &0.26&4.93&5.19&$17.03\%$ & 0.26 &0.59&\textbf{0.85} & $4.11 \times$ / $2.15\times$ / $6.10\times$\\
       
        & \multirow{1}{*}{256}  & $16M$ & 1.37 & 19.32 &20.69 &0.75 & 7.18 & 7.93 &/&/&/&/&0.02 & 3.30 & \textbf{3.32}& $6.22\times$ / $2.38\times$\\
        
        & \multirow{1}{*}{512}  &  $131M$ & 2.30 & 121.49&123.47 & 1.03 & 103.86&104.89&/&/&/&/&0.30 &9.93 & \textbf{10.23}  & $ 11.85\times$ / $ 10.25\times$\\
       
       \bottomrule
    \end{tabular}    
    }
    \label{tab:time}
 \end{table*}
\begin{figure*}
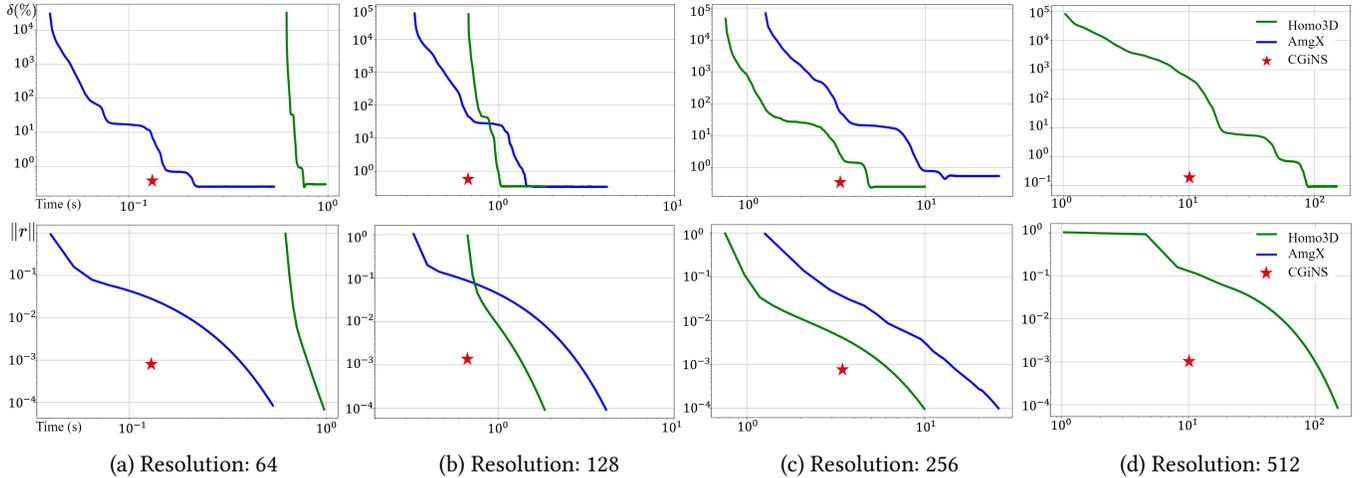

    \centering
    \includegraphics[width=\linewidth]{NMI/images/ErrorConvergenceMap.svg.pdf}
    \includegraphics[width=\linewidth]{NMI/images/ResConvergenceMap.svg.pdf}
    \leftline{ 
    \hspace{0.07\linewidth}
        (a) Resolution: 64\hspace{0.115\linewidth}
        (b) Resolution: 128\hspace{0.115\linewidth}
        (c) Resolution: 256\hspace{0.115\linewidth}
        (d) Resolution: 512}
   \caption{\revision{\textbf{Accuracy–runtime and residual–runtime comparisons across resolutions.} We compare \name (red star) with two GPU-based solvers—AmgX (blue) and Homo3D (green)—on voxel grids of size $64^3$, $128^3$, $256^3$, and $512^3$. \textbf{Top row:} Relative error of the effective elasticity tensor $\delta(\%)$ versus wall-clock time, illustrating the accuracy–runtime tradeoff. \textbf{Bottom row:} Residual norm $\Vert r \Vert$ versus time, showing convergence behavior. All plots use logarithmic axes.}}
    \label{fig:convergence-map}
\end{figure*}

As shown in \cref{tab:time}, \name achieves material property errors below $1\%$ across all tested resolutions, while offering $2\times$ to $10\times$ speedups relative to AmgX and Homo3D. \cref{fig:convergence-map} further reveals that although traditional solvers can continue reducing the residual norm $\Vert r \Vert$ through extended iterations, such improvements have negligible effect on the computed effective elasticity tensor $C^H$. This is because the mapping from displacement field $u$ to $C^H$ involves global energy integration and dot-product aggregation, which inherently smooth out local numerical perturbations and limit the practical benefit of over-solving the linear system.
This phenomenon has also been documented in the numerical homogenization literature~\cite{schneider2021review,willot2015fourier}, where errors in $C^H$ below $1\%$ are widely accepted as sufficient for engineering applications. Together, these results confirm that \name achieves not only substantial acceleration over classical solvers, but also converges to the necessary precision regime required for downstream material property prediction.

\revision{
\paragraph{Numerical–neural hybrid solver}
We also evaluate GNN-AMG~\cite{luz2020learning}, which delegates construction of the algebraic-multigrid \emph{prolongation operator} $P$ to a message-passing GNN; as in the original design, we keep the AMG pipeline unchanged and replace only the values of $P$ with the network’s predictions. Because the method requires an explicit graph for $K$ and predicts all candidate entries of $P$, in our 3D periodic-lattice setting the maximum feasible resolution is limited to $128^3$ by GPU memory and runtime. A run is deemed \emph{successful} if the linear solve reaches the prescribed tolerance.}

\revision{As summarized in \cref{tab:time}, on \textit{TPMS} GNN-AMG achieves the highest success rate among our three datasets, and its solve time is lower than AmgX. In contrast, on \textit{PSL} and \textit{Truss}, the success rate is low and solve times exceed AmgX. These observations suggest that learning $P$ directly is more effective on simpler geometries, whereas for complex and highly variable geometries it often fails to converge.
}

\subsection{PCG-Informed learning: joint optimization with embedded PCG refinement}
\begin{table}[t]
    \centering
   \caption{\textbf{Effect of PCG iteration count on accuracy, residual, and efficiency.} We evaluate the influence of different PCG smoothing step counts on prediction quality and runtime at resolution $n=64$ using the Truss dataset. As the number of iterations increases, the relative elastic tensor error ($\delta$), and residual norm ($\Vert r\Vert_{Mean}$). A small number of iterations (e.g., 10) is sufficient to achieve high accuracy with moderate additional runtime.}
    \scalebox{0.93}{
        \begin{tabular}{crrrlr}
            \toprule
             \textbf{ iteration } &  $\bm{\delta_{Max} }$ & $\bm{\delta_{Min}}$ &$\bm{\delta_{Mean}}$ &$\bm{||r||_{Mean}}$ & \textbf{Time (ms)} \\
             \midrule
             0 &$191.43\%$&$8.09\%$&$52.63\%$ & $7.18 \times 10 ^{-2}$ &80.33 \\
             1 &$30.14\%$&$1.33\%$ & $8.64\%$ & $2.14 \times 10 ^{-2}$ &133.30\\
             4 &$16.94\%$& $0.50\%$ & $4.18\%$ & $1.24 \times 10 ^{-2}$ &148.30\\
             8 & $0.91\%$ & $0.15\%$ & $0.51\%$ & $3.38\times 10 ^{-3}$& 158.27\\
             \textbf{10} & $\bm{0.70\%}$ & $\bm{0.11}\%$ & $\bm{0.50\%}$ & $\bm{3.34\times 10 ^{-3}}$& \textbf{158.91}\\
           \bottomrule
        \end{tabular}
        }
    
    \label{tab:smoother}
\end{table}

\revision{\paragraph{Joint Training with PCG}
Accurately capturing displacement fields in microstructures with complex topology presents a significant challenge for neural solvers. To address this, \name integrates a lightweight PCG solver directly into the decoder and trains the entire architecture in a solver-in-the-loop fashion. Unlike post-training numerical corrections, this joint optimization ensures that the network produces intermediate predictions already compatible with the PCG. As a result, only a few refinement steps are required to suppress residuals and recover physically meaningful solutions.}

\paragraph{Default iteration settings.}
To balance computational efficiency and predictive accuracy, we empirically determine the number of PCG iterations based on dataset complexity. For TPMS, which exhibits smooth and regular deformation patterns, 4 iterations are sufficient. PSL, with moderate geometric complexity, requires 8 iterations. For the highly irregular and discontinuous Truss dataset, 10 iterations are used.

These settings are chosen based on ablation experiments (\cref{tab:smoother}), which show diminishing returns in accuracy beyond these values.  For example, on the Truss dataset, 10 PCG steps already reduce the residual norm to $3.34 \times 10^{-3}$ and the elastic tensor error to $0.50\%$, indicating that further refinement offers marginal benefits relative to the additional cost.

\begin{figure}
    \centering
    \includegraphics[width=\linewidth]{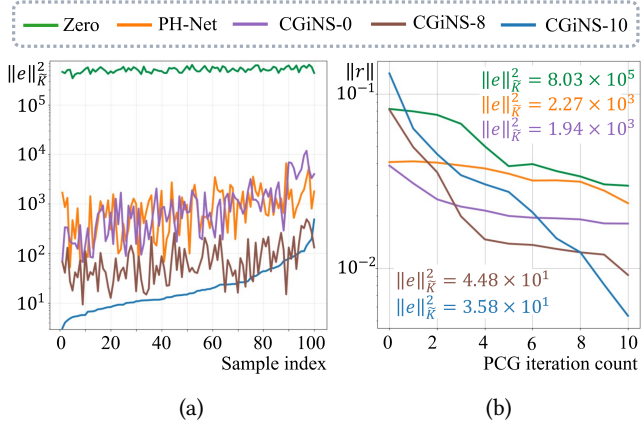}
    \leftline{ 
        \hspace{0.25\linewidth}
        (a) \hspace{0.42\linewidth}
        (b)}
    \caption{\textbf{Impact of PCG-coupled training on projection energy and convergence behavior.} (a) Initial projection energy $\Vert e\Vert_{\tilde{K}}^2$ computed on 100 microstructures using four initialization methods: Zero (classical solver baseline), PH-Net, \name-0 (trained without PCG), \name-8, and \name-10. Samples are sorted in ascending order based on \name-10 energy. (b) PCG residual norm $\Vert r\Vert$ over 10 iterations for a representative sample. \name-10 consistently achieves the lowest initial projection energy and fastest convergence, confirming the effectiveness of joint network–solver training.}
    \label{fig:smooth-map}
\end{figure}
\paragraph{Projection energy: quantifying initialization quality.}\revision{In the preceding sections we leveraged embedded PCG primarily as a lightweight refinement to suppress the network’s spectral bias. Here we adopt a complementary view of their interplay: rather than merely separating high and low frequency errors, the neural network can serve PCG by providing a high-quality initial guess that is already well aligned with the geometry of the preconditioned system. When $u_0$ lies mostly in favorable eigendirections, PCG needs only a few iterations to remove a small, well-conditioned residue. To make this notion precise, we quantify initialization quality with a simple \textit{projection energy metric}.}

To evaluate how well the predicted displacement field aligns with the dynamics of the numerical solver, we adopt the projection energy $\Vert e \Vert_{\tilde{K}}^2$ as a diagnostic metric. This metric quantifies the spectral compatibility between the network output and the eigenspace of the preconditioned stiffness matrix $\tilde{K}$, which directly impacts the convergence speed of iterative solvers like PCG.

Let $u_0$ denote the initial displacement guess (e.g., from a neural network), and $u^\ast$ the converged solution to the linear system. The error is defined as $e = u_0 - u^\ast$. 
The preconditioned stiffness matrix is $\tilde{K} = D^{-1}K$, where $D$ is the preconditioner. 
The projection energy is computed as:
\begin{equation}
\Vert e \Vert_{\tilde{K}}^2 = e^\top \tilde{K} e.
\label{eq:proj_energy}
\end{equation}

We now decompose the error $e$ into the eigenbasis of $\tilde{K}$. Suppose $\tilde{K}$ has an orthonormal set of eigenvectors ${\phi_i}$ with corresponding eigenvalues ${\lambda_i}$, such that:
\begin{equation}
\tilde{K} \phi_i = \lambda_i \phi_i.
\end{equation}
Then the error vector $e$ can be expanded as:
\begin{equation}
e = \sum_{i=1}^n \alpha_i \phi_i,
\end{equation}
where $\alpha_i = \phi_i^\top e$ are the coefficients of $e$ projected onto each eigenmode.

Substituting into \cref{eq:proj_energy}, the projection energy becomes:
\begin{equation}
\Vert e \Vert_{\tilde{K}}^2 = \sum_{i=1}^n \lambda_i \alpha_i^2.
\end{equation}

This form reveals that the total projection energy is a weighted sum of the error components, where each component is scaled by the corresponding eigenvalue $\lambda_i$. 
\revision{In PCG, the convergence rate depends strongly on the spectral distribution of the preconditioned system. Error components aligned with eigen-directions of higher energy (large $\lambda$) are typically harder to eliminate, meaning that if the initial solution carries large projection weights in these directions, convergence will be slow. Conversely, components with lower energy (small $\lambda$) are reduced more quickly, so a good initialization is one that minimizes the error energy in the high-$\lambda$ directions. In this sense, the projection energy provides a direct measure of initialization quality and its impact on subsequent PCG efficiency.}

We evaluate four initialization strategies: (1) zero initialization (as in conventional PCG solvers), (2) PH-Net predictions, and \name predictions trained with (3) 8 and (4) 10 embedded PCG steps. As shown in \cref{fig:smooth-map}, only the solver-aware training of \name produces initializations with minimal projection energy, consistently achieving the fastest convergence across all scenarios. 
This highlights the effectiveness of joint training in aligning learned representations with solver dynamics. Formal derivations of the projection energy metric are provided in Supplementary.

\subsection{Ablation study}
\label{Ablation-Study}
\paragraph{Effect of multi-level prediction architecture.}
Conventional neural solvers for microstructure modeling typically rely on single-level architectures, which often struggle to resolve fine-grained features in geometrically complex structures. To address this, \name adopts a hierarchical multi-level framework inspired by multigrid methods, where progressively refined displacement fields are predicted across a cascade of resolution levels.

We evaluated the impact of prediction depth on the Truss dataset at resolution $n=64$. As shown in \cref{tab:Layer number}, increasing the number of levels substantially improves accuracy with minimal runtime cost. Moving from a single-level model to two levels reduces the mean elasticity tensor error ($\delta_{\mathrm{Mean}}$) from $2.97\%$ to $0.34\%$, with only a modest increase in time and parameters. A third level provides a marginal improvement to $0.29\%$, indicating diminishing returns.

Based on this analysis, we select two levels for $n=64$ and $n=128$, and three levels for $n=256$ and $n=512$ to strike a balance between accuracy and computational overhead. This architectural choice plays a key role in enabling stable, high-fidelity predictions across diverse resolutions.
\begin{table}
    \centering
    \caption{\textbf{Impact of multi-level prediction depth on accuracy, residuals, and runtime.} We ablate the number of multi-resolution prediction levels in \name using the Truss dataset at resolution $n=64$. Increasing levels from 1 to 3 substantially reduces both the mean elastic tensor error ($\delta_{\mathrm{Mean}}$) and residual norm ($\Vert r\Vert$), while incurring moderate increases in inference time and parameter count. Two levels offer the best trade-off between accuracy and computational cost.}

    \scalebox{.9}{
        \begin{tabular}{crrcc}
            \toprule
             \textbf{Level} & \multicolumn{1}{c}{$\bm{\delta_{Mean}}$ }& \multicolumn{1}{c}{$\bm{\Vert r\Vert}$} & \multicolumn{1}{c}{\textbf{Time (ms)}} & \multicolumn{1}{c}{\textbf{Parameter (M)}} \\
             \midrule
             1 & $2.97\%$ &$1.04\times10^{-2}$ & $140.05$ &$7$\\
             \textbf{2} & $\bm{0.34\%}$ &$\bm{3.13\times10^{-3}}$ & $158.27$ & $12$\\
             3 & $0.29\%$ &$2.75\times10^{-3}$ & $166.05$ &$18$\\
           \bottomrule
        \end{tabular}
        }
    
    \label{tab:Layer number}
\end{table}

\begin{table}[t]
    \centering
    \caption{\textbf{Ablation study.} We conducted ablation experiments on TPMS and Truss with $n=64$, where each experiment contained 1,000 random microstructures with random materials. We report the average error $\delta_{Mean}$ and average residual $\Vert r\Vert_{Mean}$. 
    The second row shows results of \name without the CG Solver: \name (WO-CS) .
    The third row presents \name without the global displacement constraint: \name (WO-GD), where Dirichlet boundary conditions were used.
    The fourth row represents \name without the addition of soft voxels: \name (WO-SV).
    The fifth row shows \name without the Peri-mapping data structure: \name (WO-PM).}
    \scalebox{0.95}{
        \begin{tabular}{lrr|rr}
            \toprule
             \multirow{2}{*} {\textbf{Network} } & \multicolumn{2}{c}{\textbf{TPMS}} & \multicolumn{2}{c}{\textbf{Truss}} \\
              & $\bm{\delta_{Mean}}$ & \multicolumn{1}{c}{$\bm{\Vert r\Vert_{Mean}}$} & $\bm{\delta_{Mean}}$ & \multicolumn{1}{c}{$\bm{\Vert r\Vert_{Mean}}$ }  \\
             \midrule
             \name          &$\bm{0.05\%}$  &$\bm{1.48\times10^{-3}}$ & $\bm{0.34\%}$ &$\bm{3.38\times10^{-3}}$  \\
             \name (WO-CS)  &$0.11\%$  &$2.31\times10^{-3}$    &$52.63\%$ &$7.18\times10^{-2}$\\
             \name (WO-GD)  &$0.59\%$   &$4.43\times10^{-3}$    &$39.90\%$ & $5.18\times10^{-2}$\\
             \name (WO-SV)  &$0.57\%$   &$4.43\times10^{-3}$    &$11.41\%$ &$2.52\times10^{-2}$\\
             \name (WO-PM)  &$0.75\%$  & $5.08\times10^{-3}$   &$9.06\%$ & $2.02\times10^{-2}$\\
            \bottomrule
        \end{tabular}
        }
    
    \label{tab:Ablation}
 \end{table}



\paragraph{PCG iterations.}
An essential component of \name is the embedded numerical solver, which ensures that the predicted displacement fields satisfy the equilibrium constraints governed by PDEs. To validate its necessity, we conducted an ablation experiment in which the PCG solver was entirely removed from the inference pipeline. That is, the network outputs were used directly as displacement fields without any iterative correction.

As shown in \cref{tab:Ablation}, \name (WO-CS) yielded a relative error of only $0.11\%$ on TPMS, a dataset characterized by smooth geometries and strong structural regularity. However, when tested on the more topologically complex Truss dataset, the error dramatically increased to $52.63\%$, highlighting severe prediction instability. This result confirms that the numerical solver is not merely an auxiliary refinement tool but a necessary component for achieving convergence, especially irregular microstructures. Even with a strong initial guess provided by the network, convergence cannot be guaranteed without the PCG module.

These findings reinforce our design choice: \name must be viewed as a hybrid framework in which the neural network provides a high-quality initialization, and the embedded solver ensures physical consistency and convergence.

\paragraph{Global displacement constraint.}
Accurate PDE solutions require the enforcement of Dirichlet boundary conditions, often implemented in homogenization by fixing the displacement of a designated anchor node. Existing network-based approaches, such as PH-Net and Label-Free, typically fulfill this requirement by hard-coding the first indexed node to zero displacement after mesh preprocessing. While straightforward, this approach assumes that the first node corresponds to a geometrically meaningful and spatially consistent anchor -- which holds only for highly regular geometries like TPMS.

However, for complex microstructures such as Truss, no consistent or meaningful anchor can be inferred across samples, and this anchor-node heuristic becomes unreliable. In \name, we address this limitation by removing the dependence on arbitrary node selection and introducing a global displacement constraint: we enforce that the mean displacement over all nodes equals zero. This formulation inherently satisfies translation invariance and avoids structural assumptions, making it more robust across diverse datasets.

To quantify the effectiveness of this design, we evaluate \name (WO-GD), a variant in which anchor-based constraints are reinstated. Specifically, we adopt the same strategy as PH-Net and Label-Free -- fixing the displacement of the first node in the loss function. As shown in \cref{tab:Ablation}, this modification leads to a marked degradation in accuracy, particularly on Truss, demonstrating that the network alone cannot infer a physically consistent anchor. The global constraint is thus not merely a convenience -- it is essential for stable and transferable learning. 
The global displacement constraint replaces the reliance on fixed anchor nodes -- common in prior homogenization models -- with a translation-invariant formulation, ensuring stability and transferability across diverse microstructural configurations. This addresses a key limitation of traditional Dirichlet-boundary-based models, which often fail to generalize reliably across irregular topologies.

\paragraph{Soft voxel padding.}
To support high-resolution microstructure prediction under tight memory constraints, \name adopts a sparse tensor representation. While efficient, this sparsity makes it harder to capture fine-grained geometric details—particularly near material boundaries where voxel occupancy changes abruptly. To address this, we introduce soft voxel padding: a lightweight mechanism that surrounds solid voxels with low-stiffness buffer zones. These soft voxels have negligible mechanical impact but improve the representation of boundary geometry, enabling convolutional kernels to better resolve shape transitions.

As shown in \cref{tab:Ablation}, removing this component (\name (WO-SV)) results in a consistent drop in prediction accuracy, confirming its role in maintaining structural fidelity -- particularly in topologically intricate regions.


\paragraph{Peri-Mapping data structure.}
Accurate homogenization relies on enforcing periodic boundary conditions (PBCs), which are essential for capturing cross-scale mechanical behavior. However, existing learning-based methods typically implement PBCs as soft constraints -- penalizing mismatches across boundaries in the loss function. This weak enforcement often fails to preserve true periodicity, especially in complex geometries.

To address this, \name introduces a Peri-mapping data structure that explicitly links corresponding node pairs on opposing boundaries during convolution operations. Further implementation details are provided in~\cref{method:Peri-mapping}.

Ablation results on the Truss dataset (\cref{tab:Ablation}) show that removing Peri-mapping (\name (WO-PM)) increases the mean error to $9.06\%$, confirming that strict periodicity enforcement is critical for accurate and stable predictions across structurally complex microstructures.

\revision{
\subsection{Multiphase homogenization case}
\begin{table}[t]
    \centering
    \caption{\textbf{Multiphase homogenization with \name on the Truss dataset.} Results for test cases with varying numbers of phases (2, 3, 5, and INF, where INF denotes fully random phase assignment). We recorded the relative error of the computed solution with respect to the reference solution ($\delta_{Mean}$), and the residual norm of the linear system ($\Vert r\Vert_{Mean}$).}
    \scalebox{.95}{
        \begin{tabular}{lcccc}
            \toprule
             \multirow{2}{*} {\textbf{ Metrics} } & \multicolumn{4}{c}{\textbf{Number of Phases}}\\
             & \multicolumn{1}{c}{\textbf{2}}
             & \multicolumn{1}{c}{\textbf{3}}
             & \multicolumn{1}{c}{\textbf{5}}
             & \multicolumn{1}{c}{\textbf{INF}}\\
             \midrule
             $\delta_{Mean}$ & $0.33\%$&$0.65\%$&$0.90\%$&$1.01\%$\\
             $\Vert r\Vert_{Mean}$ & $2.11\times10^{-3}$ &$5.52\times10^{-3}$&$4.83\times10^{-3}$&$6.09\times10^{-3}$ \\
            \bottomrule
        \end{tabular}    
        }
    \label{tab:A-Phase}
 \end{table}

In line with practical manufacturing, our main experiments emphasize single-phase lattices, which dominate current fabrication tasks, have mature design–for–manufacture pipelines, and are standard building blocks for lightweighting, energy absorption, and biomedical implants; periodic unit-cell single-phase lattices of this kind are ubiquitous in industrial workflows. }

\revision{
Importantly, \name encodes material parameters at each voxel and thus naturally extends to multiphase homogenization without modifying the overall pipeline or requiring retraining. 
To illustrate this capability, we constructed an dataset of 10,000 Truss microstructures in which each voxel can take a distinct material: 2-phase, 3-phase, 5-phase, and an “INF” case with randomly phased assignments (2,500 samples per category). Materials are non-dimensionalized with $E=1$ and $\nu \in\left[ 0.2,0.4\right]$. We conducted additional tests on benchmark datasets, using \name pretrained on single-phase materials (Table~\ref{tab:material}'s pretrained model).}

\revision{As summarized in Table~\ref{tab:A-Phase}, mean residual norms for the linear solves are on the order of $10^{-3}$, while the mean relative error of the elasticity tensor remains below $1\%$ for 2-phase ($0.33\%$), 3-phase ($0.65\%$), and 5-phase ($0.90\%$) cases; the INF setting yields $1.01\%$, a slight increase attributable to strong material jumps across neighboring voxels. These results indicate that \name  maintains high accuracy for multiphase homogenization and is readily extensible beyond single-phase microstructures.
}

\section{Discussion}
\label{sec:discussion}

\revision{This work introduces \name, which tightly couples geometry-aware sparse convolutions, strict periodic encoding, and a lightweight PCG refinement to jointly achieve accuracy, stability, and efficiency for high-resolution periodic microstructure homogenization.  Unlike end-to-end regressors or approaches with weak physical priors, \name encodes periodic boundaries and the global displacement constraint explicitly in the network, and then aligns the prediction with the dynamics of a numerical solver via a few PCG steps during inference.  This coupling yields two immediate benefits: (i) errors of the effective elasticity tensor remain below $1\%$ with residuals on the order of $1\times10^{-3}$ across complex topologies;  and (ii) end-to-end speedups of $2\times$ to $10\times$ over GPU multigrid solvers while scaling to $512^3$resolutions, where hybrid methods often suffer from memory or convergence issues.}

\revision{Across three representative datasets (TPMS, PSL, and Truss), \name consistently reduces effective-tensor error and improves numerical stability compared with 3D-CNN, Label-Free, PH-Net, and other baselines.  The gains are most pronounced on geometrically irregular cases (PSL, Truss) with strong high-frequency content in the displacement field, where baselines frequently exceed $10\%$ error or become unstable, whereas \name maintains $\textless1\%$ error and lower residuals.  These results indicate that a tightly coupled “Neural + Numerical” pipeline mitigates spectral bias and preserves physical consistency.}

\revision{Across resolutions from $64^3$ to $512^3$, \name achieves $2\times$ to $10\times$ end-to-end speedups over GPU multigrid solvers (AmgX/Homo3D) at matched accuracy thresholds, and at $512^3$ the overall wall-clock improvement reaches $8.6\times$ to $11.9\times$ on TPMS/PSL/Truss. Notably, once the homogenized tensor error is around $1\%$, further driving the linear-system residual lower with classical solvers yields negligible gains in $C^H$, which explains \name’s favorable accuracy–time trade-off: a few physically consistent PCG steps suffice to reach industrial-grade accuracy. At the same time, results with learned numerical hybrids—e.g., learning the AMG P-operator with GNNs—reveal that a plain "network $+$ solver" recipe is not enough in complex settings. While such methods can accelerate regular geometries (e.g., TPMS), they degrade in success rate and efficiency on irregular, high-contrast cases (PSL, Truss). The core issue is a lack of geometry-aware inductive bias: without explicitly encoding periodicity, boundary handling, and global constraints, the learned transfer/relaxation operators become fragile under geometric variability. \name addresses this gap by embedding geometry and physics directly into the learning system—via geometry-aware sparse convolutions, strict periodic encoding, and a global displacement constraint—then aligning the prediction with a lightweight in-the-loop PCG. This combination preserves multigrid-like cross-scale structure while remaining robust on complex topologies, delivering both the stability that learned hybrids struggle to maintain and the speedups that classical solvers alone cannot match at comparable accuracy.}

\revision{Ablations clarify the role and limits of each component:
(1) PCG-in-the-loop. Injecting a small number of PCG steps after the network prediction markedly reduces residuals and restores physically interpretable solutions; benefits plateau around ~10 steps. Removing PCG causes sharp error increases on complex topologies (Truss), showing it is essential for convergence and stability rather than a cosmetic refinement.
(2) Multi-level prediction. Increasing from one to two levels significantly lowers error and residual; a third level shows diminishing returns, underscoring the need for cross-scale features for high-fidelity displacement fields.
(3) Global displacement constraint. Enforcing a zero-mean displacement in lieu of heuristic Dirichlet anchors improves consistency and transferability on complex meshes, avoiding anchor ambiguity.
(4) Soft-voxel boundary buffering and peri-mapping for periodicity improve boundary-geometry fidelity under sparse representations and preserve periodic consistency; ablating either degrades errors and residuals in a systematic way.}

\revision{Regarding material generalization and multi-phase extensions, \name encodes isotropic matrix parameters explicitly as inputs, enabling stable predictions across material settings. Even for 2/3/5-phase and voxel-level random phase assignments (INF), errors of the effective stiffness remain around $1\%$. This suggests that a self-supervised energy-minimization loss combined with PCG-in-the-loop training preserves physical consistency and numerical robustness across the coupled geometry–material space.}

\revision{\paragraph{Limitations and outlook.} This work targets linear elasticity with periodic boundaries and a global displacement constraint (Dirichlet-equivalent). More general Neumann/mixed boundaries, nonlinear materials, and contact are not yet covered. The current $512^3$ ceiling is mainly due to memory and implementation limits; future improvements include mixed precision, domain decomposition, and more memory-efficient operators. A key direction is cross-resolution prediction: train at a single “base” resolution and deploy zero or few-shot across other resolutions using scale-conditioned encodings plus multigrid-consistent restriction and prolongation to preserve invariants. Another avenue is replacing fixed preconditioners with learnable yet stability-controlled modules, and extending the framework to broader PDEs. The overarching lesson remains: maintain a physically consistent feedback between the network and the numerical iteration to reach industrial-grade accuracy with minimal extra solves.}
\section{Methods}
\label{sec:method}

In this section, we detail the prediction framework for the linear elastic mechanical properties of periodic microstructures, as implemented in \name. An overview of the key components is provided in \cref{fig:Pipeline}.

\subsection{The input and output of \name}
\paragraph{Input: Material-Structure encoding}
The effective elastic properties of a microstructure depend on both its geometric topology and the mechanical properties of the base material. Accordingly, \name takes a combined encoding of structure and material as input.

Each periodic microstructure $\Omega$ is discretized into a voxel grid at resolution $n$~\cite{dong2019149}. The geometric information is encoded using a Fourier feature network $\mathcal{F}(\Omega)$~\cite{tancik2020fourier}, which projects node positions into a higher-dimensional space. Fourier features are employed to mitigate the spectral bias commonly observed in neural networks, thus enhancing the model’s ability to learn high-frequency variations in the displacement field.

The mechanical behavior of the base material is characterized by the elasticity tensor $C^b \in \mathbb{R}^{6 \times 6}$, determined by its Young’s modulus $E$ and Poisson’s ratio $\nu$ (i.e., $C^b = \mathcal{M}(E, \nu)$; see Supplementary Sec.1 for details). 
To improve geometric detail representation near boundaries, soft voxels are introduced with a significantly reduced modulus, where $C^b_{\mathrm{soft}} = \mathcal{M}(10^{-6} \times E, \nu)$. This design ensures accurate boundary modeling without introducing artificial stiffness effects that could distort the mechanical response.

Let $\Omega_e$ and $\Omega_s$ denote the sets of solid and soft voxels, respectively. Using $\star$ to represent matrix filling and $\oplus$ for matrix concatenation, the final network input $\mathcal{X}$ is constructed as: 
\begin{equation} \mathcal{X} = (\Omega_e \star C^b) \oplus \mathcal{F}(\Omega_e) + (\Omega_s \star C^b_{\mathrm{soft}}) \oplus \mathcal{F}(\Omega_s). 
\label{eq:4-1} 
\end{equation}




\paragraph{Output: Node displacement}
The output of \name is the nodal displacement field $u \in \mathbb{R}^{6 \times 3}$, where each row corresponds to the displacement components in three Cartesian directions under six independent loading cases.

\subsection{The components of \name}
\label{subsec:4.2}

\paragraph{Prediction Network}

The core prediction network in \name follows an encoder-decoder structure. The input $\mathcal{X}$ is first mapped into feature representation $\mathcal{Q}$ via an encoder. A U-Net~\cite{ronneberger2015u} architecture processes $\mathcal{Q}$, generating an intermediate displacement latent code $\mathcal{Z}$. Finally, a decoder maps $\mathcal{Z}$ to the nodal displacement field $u$.


Periodic boundary conditions are enforced throughout convolutional operations using the Peri-mapping data structure, ensuring continuity across periodic interfaces in every convolutional layer. To enhance computational efficiency, the entire framework is implemented with sparse matrices, utilizing the Minkowski engine~\cite{choy20194d} for efficient sparse convolution operations.

\paragraph{Multi-level Frame}
To improve predictive accuracy and scalability, \name incorporates a multi-level prediction framework inspired by multigrid methods. The finest resolution microstructure domain $\Omega$ serves as the input to the first-level, $\mathcal{Q}^1$. For an $L$-level network, the input is recursively coarsened to generate a hierarchy of feature maps $\mathcal{Q}^1, \mathcal{Q}^2, \ldots, \mathcal{Q}^L$.

Information transfer between levels is achieved using trilinear interpolation and its transpose, acting as prolongation ($I_{l+1}^l$) and restriction ($R^{l+1}_l$) operators, respectively. The features propagate across levels according to: $\mathcal{Q}^{l+1}=R^{l+1}_l\mathcal{Q}^{l}$ and $\mathcal{Z}^{l}=\mathcal{Z}^{l}+I_{l+1}^l\mathcal{Z}^{l+1}$. This multi-resolution structure enhances the network's ability to capture both global and local deformation patterns in complex microstructures.

\paragraph{Peri-mapping data structure}
\label{method:Peri-mapping}

\begin{figure}
    \centering
    \includegraphics[width=\linewidth]{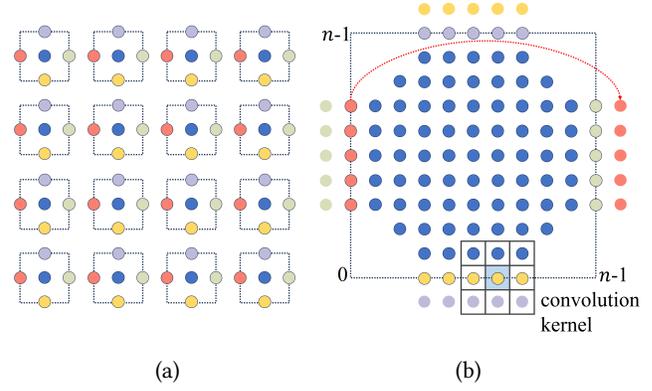}
    \leftline{ 
        \hspace{0.20\linewidth}
        (a) \hspace{0.41\linewidth}
        (b)}
    \caption{\textbf{2D illustration for Peri-mapping data structure.} 
(a) Periodic tiling of a microstructure in space, where the domain repeats periodically. (b) To emulate periodic boundary conditions within a single simulation cell, \name remaps edge nodes (e.g., red to red) across domain boundaries, allowing convolution kernels to access wrapped-around neighbors. This ensures explicit preservation of boundary continuity during feature extraction. The schematic shows a 2D example but applies equally to 3D.
}
    \label{fig:periodic-convolution}
\end{figure}

Accurate homogenization requires strict enforcement of periodic boundary conditions (PBCs), which are essential for capturing cross-scale mechanical behavior. Existing learning-based methods often impose PBCs as soft constraints in the loss function, which cannot guarantee strict periodicity, especially in complex geometries. To address this, \name introduces a Peri-mapping data structure that explicitly encodes cyclic connectivity into the network architecture (\cref{fig:periodic-convolution}).

For each pair of opposing domain boundaries—such as the left and right faces along the x-axis—Peri-mapping establishes a one-to-one correspondence between spatially aligned node positions. For example, the node at $(0, y, z)$ is mapped to $(n, y, z)$, where $n$ is the resolution along that axis. Similar mappings are constructed along the $y$ and $z$-axes, ensuring that all periodic boundary pairs are covered. This index mapping is used in sparse convolution, enabling each node near a boundary to directly access features from its periodic counterpart. As a result, periodicity is enforced directly within the computation graph, eliminating the need for auxiliary loss terms and ensuring consistent boundary behavior by design.

\paragraph{Soft voxel}
\begin{figure}
    \centering
    \includegraphics[width=.8\linewidth]{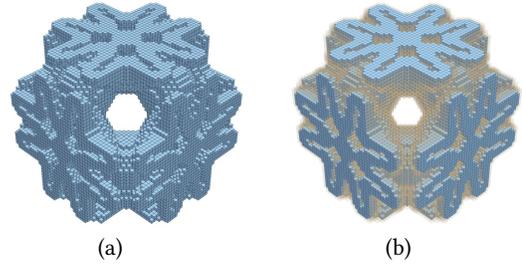}
        \leftline{ 
        \hspace{0.22\linewidth}
        (a) \hspace{0.39\linewidth}
        (b)}
    \caption{\textbf{Soft voxel padding for boundary-aware representation.} Left: Original voxelized microstructure. Right: Soft voxels (shown in translucent beige) are added around the boundary to preserve geometric context and enhance convolutional feature continuity near edges, while incurring minimal memory overhead.}
    \label{fig:soft-voxel}
\end{figure}
Predicting displacement fields on sparse voxel grids reduces memory consumption but introduces challenges in accurately capturing geometric detail near complex material boundaries. In particular, sparsity can cause spatially close but unconnected nodes to be mistakenly treated as adjacent, leading to artifacts in predicted displacements.

To address these issues, \name introduces a soft voxel padding strategy (\cref{fig:soft-voxel}). A thin layer of soft voxels is added around all solid voxels. These soft voxels share the same Poisson’s ratio $\nu_{\mathrm{soft}}$ as the base material but have a significantly reduced Young’s modulus $E_{\mathrm{soft}} = 10^{-6} \times E$, ensuring minimal influence on physical behavior while enhancing boundary continuity. This strategy improves the model’s ability to learn displacement fields near interfaces without introducing artificial stiffness.

\paragraph{PCG solver.}
To improve numerical stability and ensure boundary condition consistency, the predicted displacement fields $u$ are globally translated by subtracting their mean value $\bar{u}$, yielding $u - \bar{u}$. This enforces a zero-mean displacement constraint, equivalent to applying global Dirichlet-type boundary conditions in homogenization.

To further reduce residuals and suppress high-frequency prediction errors, we apply a $t$-step smoothing process using a PCG solver. For low-resolution microstructures (e.g., $n=64$ or $n=128$), we employ an algebraic PCG method based on explicit sparse matrix assembly of the global stiffness matrix $K$ and right-hand side vector $f$. This strategy allows for rapid convergence within a few iterations, leveraging standard linear algebra routines on small matrices.

For high-resolution structures (e.g., $n=256$ or $n=512$), assembling the full stiffness matrix becomes prohibitively memory-intensive. Instead, we adopt a geometry-based PCG implementation that avoids explicit matrix construction by iteratively computing stencil operations over local 27-node neighborhoods. This matrix-free formulation significantly reduces memory consumption while maintaining numerical consistency.
Further implementation details can be found in Supplementary.


\subsection{Loss function}
\paragraph{Energy loss function} 
\name adopts a self-supervised training strategy based on the principle of minimum potential energy, eliminating the need for labeled datasets and enhancing model generalization. In linear elastic systems, the equilibrium state minimizes the total potential energy functional, with the first variation vanishing at equilibrium~\cite{reddy2017energy}.
To enforce this physical principle during training, \name constructs an energy-driven loss function.

\paragraph{Global boundary conditions} 

As \cref{eq:2-2} admits non-unique solutions without boundary constraints, appropriate boundary conditions are required. While Dirichlet boundary conditions are commonly used in numerical solvers and learning-based models, directly fixing a random node at zero displacement can lead to instability in datasets with complex topologies, as the network struggles to infer the anchor node's location and its relation to neighboring nodes.

To address this, \name imposes a global displacement constraint that is more amenable to neural network learning: the sum of displacements in each orthogonal direction is enforced to be zero for each microstructure. This constraint is equivalent to fixing rigid body motions while avoiding dependence on specific node selections.

Formally, let $k(K)$ denote the dimensionality of the stiffness matrix $K$, $D(K)$ the \revision{Jacobi} preconditioner used in the PCG solver, and $P_t(\cdot)$ the result after $t$ iterations of PCG. Let $u$ denote the predicted displacement field and $\bar{u}$ its mean over all nodes. 
To enforce translation-invariant displacement predictions, we define $\widetilde{u} = u - \bar{u}$.
The energy-based loss function at the finest resolution is then formulated as: 
\begin{equation} 
\begin{split} 
\mathcal{L} = &\frac{1}{k} \sum_{(\mathcal{M},\Omega)} \left| \frac{1}{2} P_t(\widetilde{u})^T D(K)^{-1} K P_t(\widetilde{u}) \right. \left. - P_t(\widetilde{u})^T D(K)^{-1} f \right|_1 +\\
&\lambda | \bar{u} |, 
\end{split} 
\label{eq:MPE} 
\end{equation} 
where $\lambda$ controls the penalty on the mean displacement to enforce the global constraint.

\paragraph{Multi-level energy loss function} 
Given the hierarchical structure of \name, the energy loss is computed not only at the finest resolution but also across all levels of the multi-level architecture to accelerate convergence and improve prediction accuracy.

Let $I_{l+1}^l$ and $R_l^{l+1}$ denote the prolongation and restriction operators between levels $l$ and $l+1$, respectively. The stiffness matrices and load vectors at each level are recursively constructed as: \begin{equation} 
\begin{split} 
K^{l+1} &= R_l^{l+1} K^l I_{l+1}^l, \\
f^{l+1} &= R_l^{l+1} f^l. 
\end{split} 
\label{eq:multi-layerMPE} 
\end{equation}

The total multi-level energy loss functional is defined as: 
\begin{equation} 
\begin{split} 
\mathcal{L} =& \frac{1}{L} \sum_{l=1}^{L} \frac{1}{k^l} \sum_{(\mathcal{M},\Omega)} \left| \frac{1}{2} P_t(\widetilde{u^l})^T D(K^l)^{-1} K^l P_t(\widetilde{u^l}) -  P_t(\widetilde{u^l})^T D(K^l)^{-1} f^l \right|_1 \\ &+ \lambda | \bar{u}^l |, 
\end{split} 
\label{eq:MPE-level} 
\end{equation} 
where $u^l$ and $\bar{u}^l$ denote the displacement and mean displacement at level $l$.

This multilevel energy loss function accelerates learning at coarse levels and stabilizes convergence at finer resolutions, leading to improved training efficiency and predictive robustness.


\bibliographystyle{ACM-Reference-Format}
\bibliography{NMI/src/ref}
\newpage
\appendix

\section{Homogenization theory}
\label{A-Homogenization}
This section provides a rigorous finite element formulation for homogenization analysis of periodic lattice structures.
$\Omega$ is a periodic lattice microstructure, whose homogenized linear elastic properties can be characterized  by the elastic tensor $C^H \in \mathbb{R}^{6 \times 6}$:
\begin{equation}
\begin{split}
C^H_{ijkl}=&\mathcal{H}_{omo}(\Omega,\mathcal{M})\\
=&\frac{1}{\left| \Omega \right|}\int \left( \overline{\epsilon}_{ij} - \epsilon_{ij}(u)\right)
            C^b
            \left( \overline{\epsilon}_{kl} - \epsilon_{kl}(u)\right)\,\mathrm{d} \Omega.
\end{split}
    \label{eq:A-1}
\end{equation}
Here, $C^H_{ijkl}$ represents the elastic tensor of the lattice microstructure with Voigt notation. $i,j,k,l \in \{1,2,3\}$, where 1, 2, and 3 represent the $x$, $y$, and $z$ directions, respectively.
$\overline{\epsilon}_{pq}$ is the prescribed macroscopic strain field, in this work, we consider the following six strain fields:
$$\overline{\epsilon}_{11}= (1,0,0,0,0,0)^T, \overline{\epsilon}_{22}=( 0,1,0,0,0,0)^T , \overline{\epsilon}_{33}=( 0,0,1,0,0,0)^T,$$
$$\overline{\epsilon}_{23}=(0,0,0,1,0,0)^T, \overline{\epsilon}_{13}=(0,0,0,0,1,0)^T, \overline{\epsilon}_{12}=(0,0,0,0,0,1)^T.$$
$\epsilon_{pq}(u)$ is the locally solved strain field, defined as: 
\begin{equation}
    \epsilon_{pq}(u)=\frac{1}{2}
    \left(\frac{\partial u_p}{\partial q}+\frac{\partial u_q}{\partial p}\right).
    \label{eq:A-2}
\end{equation}
 
Displacement field $u \in \mathbb{R}^{6 \times 3}$ is obtained by solving the mechanical PDE~ \cite{andreassen2014determine,dong2019149}, whose weak Galerkin formulation is expressed as:
\begin{equation}
    \begin{cases}
    \int_{\Omega}C^b_{ijpq}\epsilon_{ij}(v)\epsilon_{pq}(u)\,\mathrm{d} \Omega=
    \int_{\Omega}C^b_{ijpq}\epsilon_{ij}(v)\overline{\epsilon}_{pq}\,\mathrm{d} \Omega, \forall v \in \Omega, \\
    \epsilon_{ij}(x)=\epsilon_{ij}(x+t), x \in \partial \Omega.
    \end{cases}
    \label{eq:A-3}
\end{equation}
$v$ is a virtual displacement field. $C^b$ is the elasticity tensor of the microstructure base material, which is isotropic and depends on the base material's Young's modulus $E$ and Poisson's ratio $\nu$.
We define $\Theta=\frac{\nu}{1-\nu}$ and $\Phi=\frac{1-2\nu}{2(1-\nu)}$.
It can be computed as:
\begin{equation}
\begin{split}
        C^b&=\mathcal{M}(E,\nu)\\
        &=\frac{E(1-\nu)}{(1+\nu)(1-2\nu)}
        \begin{bmatrix}
        1 & \Theta&\Theta & 0 & 0 & 0 \\
        \Theta& 1 & \Theta & 0 & 0 & 0 \\
        \Theta & \Theta & 1 & 0 & 0 & 0 \\
        0 & 0 & 0 & \Phi & 0 & 0 \\
        0 & 0 & 0 & 0 & \Phi & 0 \\
        0 & 0 & 0 & 0 & 0 & \Phi
        \end{bmatrix}.
        \end{split}
    \label{eq:A-3a}
\end{equation}

We use the finite element method (FEM) to numerically solve the aforementioned equations.
In this work, eight-node hexahedral elements are employed, with natural coordinates denoted as $(\xi, \eta, \zeta)$. 
The shape function is defined as:
\begin{equation}
    N=\frac{1}{8}(1+\xi\xi_i)(1+\eta\eta_i)(1+\zeta\zeta_i).
\end{equation}

$B$ is the strain-displacement matrix, defined as $B=LN$, where $L$ is the differential operator. 
The element stiffness matrix $K_e \in \mathbb{R}^{24 \times 24}$ and the element load vector $f_e \in \mathbb{R}^{24 \times 6}$ can be defined as :
\begin{equation}
\begin{split}
  K_e &= \int_{\Omega}B^TC^bB\,\mathrm{d} \Omega,\\
  f_e &= \int_{\Omega}B^TC^b\overline{\epsilon}\,\mathrm{d} \Omega.
\end{split}
\end{equation}

After discretizing the microstructure, we assign degree-of-freedom (DOF) labels. Assemble the global stiffness matrix 
$K(\Omega,\mathcal{M})$ and global load vector 
$f(\Omega,\mathcal{M})$ by mapping elemental stiffness matrices $K_e$ and load vector $f_e$ through DOF connectivity (local-to-global node numbering). 
Solve the linear system to obtain the displacement field $u$:
\begin{equation}
    Ku=f.
\end{equation}
By establishing the relationship between the displacement field and the strain field using $B$, \cref{eq:A-2} can be rewritten as:
\begin{equation}
    \epsilon(u)=Bu,
    \label{eq:A-5}
\end{equation}
and \cref{eq:A-1} can be rewritten as:
\begin{equation}
\begin{split}
    C^H_{ijkl}&=\frac{1}{\left| \Omega \right|} \sum_{e\in\Omega} \left( \overline{\epsilon}_{ij} - \epsilon_{ij}(u)\right)
            C^b_{ijkl} \left( \overline{\epsilon}_{kl} - \epsilon_{kl}(u)\right), \\
            &=\frac{1}{|\Omega |} \sum_{e\in\Omega} ( x_0 - u)^T K_e ( x_0 - u),
\end{split}
\label{eq:A-6}
\end{equation}
where $x_0=K_e^{-1}f_e$.
\section{Dataset}
\label{A-Dataset}
\subsection{Dataset introduction}
\paragraph{TPMS (Triply Periodic Minimal Surfaces)}
Triply Periodic Minimal Surfaces (TPMS) are surfaces with vanishing mean curvature at every point, satisfying the minimal surface equation. When periodic along three orthogonal directions, they form TPMS architectures exhibiting several key properties: (1) intrinsic smoothness that promotes uniform stress distribution; (2) bi-continuous topology, allowing both solid and void phases to percolate without isolated domains; (3) hyperbolic geometry with negative Gaussian curvature, enhancing surface-to-volume ratios; and (4) mechanical anisotropy, achieving superior strength-to-weight ratios compared to conventional lattices.

TPMS structures serve as fundamental templates for multifunctional supports in additive manufacturing, metamaterials, and biomedical implants. Our TPMS dataset is generated by evaluating implicit equations (\cref{tab:A-TPMS}) and extracting isosurfaces via the Marching Cubes algorithm~\cite{lorensen1998marching}.
\begin{table}
    \centering
        \caption{\textbf{Implicit surface formulations used to generate TPMS unit cells.}}
    \scalebox{.8}{
        \begin{tabular}{ll}
            \toprule
             \textbf{Type} & \textbf{Function} $\bm{T(x,y,z)}$ \\
             \midrule
             P & $\cos x + \cos y + \cos z =c$\\
             G & $\sin x\cos y + \sin z\cos x + \sin y\cos z =c$\\
             D & $\cos x\cos y \cos z - \sin x\sin y\sin z =c$\\
             I-WP & $2(\cos x\cos y + \cos y\cos z +\cos x\cos z) - (\cos2 x + \cos 2y +\cos 2x) =c$\\
             F-RD & $4(\cos x\cos y \cos z)-(\cos 2x\cos 2y + \cos2 y\cos2 z +\cos2 x\cos2 z)=c$\\
           \bottomrule
        \end{tabular}
        }

    \label{tab:A-TPMS}
\end{table}
\paragraph{Truss Lattices}
Truss microstructures are periodic frameworks composed of interconnected truss units, designed to achieve ultra-lightweight yet mechanically robust architectures~\cite{liu2008optimum}. Optimized spatial configurations enable remarkable specific strength and stiffness, making trusses a staple in engineering design, lightweight materials, and advanced manufacturing.

The truss dataset is constructed by first generating base topologies according to~\cite{saeb2016aspects,panetta2015elastic}, then scaling the thickness of individual members to meet target volume fraction (VF) constraints, and finally converting the geometry into explicit 3D voxel models.

\paragraph{PSL (Parametric Shell Lattices)}
Parametric Shell Lattices (PSL) are advanced generative designs that leverage truss-based topologies to create complex, curved surfaces~\cite{liu2022parametric}. PSL structures offer enhanced geometric control, enabling tailored node densities, strut curvatures, and stress distributions, while maintaining manufacturability under additive manufacturing constraints.

Compared to TPMS, PSL architectures provide superior flexibility for locally tuning stiffness and achieving hierarchical material properties. The PSL dataset is generated via parametric modeling with multi-objective optimization.

\subsection{Dataset construction}
\label{A-dataset-construction}
Three microstructure-specific datasets were constructed, as summarized in \cref{tab:dataset}.
For all datasets, the base material's Young's modulus was normalized to $E=1$, and soft voxels were assigned a significantly lower stiffness, $E_{\mathrm{soft}}=10^{-6}\times E$.

To capture a broad range of material behaviors, the Poisson's ratio $\nu$ was randomly sampled from the interval $[0.2, 0.4]$, corresponding to the typical range for metallic materials~\cite{koster1961poisson}.
This stochastic sampling introduces material diversity into the dataset, supporting generalization across different base material properties.

However, for the experiments specifically comparing \name with network baseline methods, material variability was suppressed.
To ensure fair and consistent comparisons -- since baseline methods are typically trained for a fixed material configuration -- we fixed the material properties to $E=1$ and $\nu=0.3$ across all microstructures used in the comparative evaluations.

\begin{figure*}
    \centering
    \includegraphics[width=.9\linewidth]{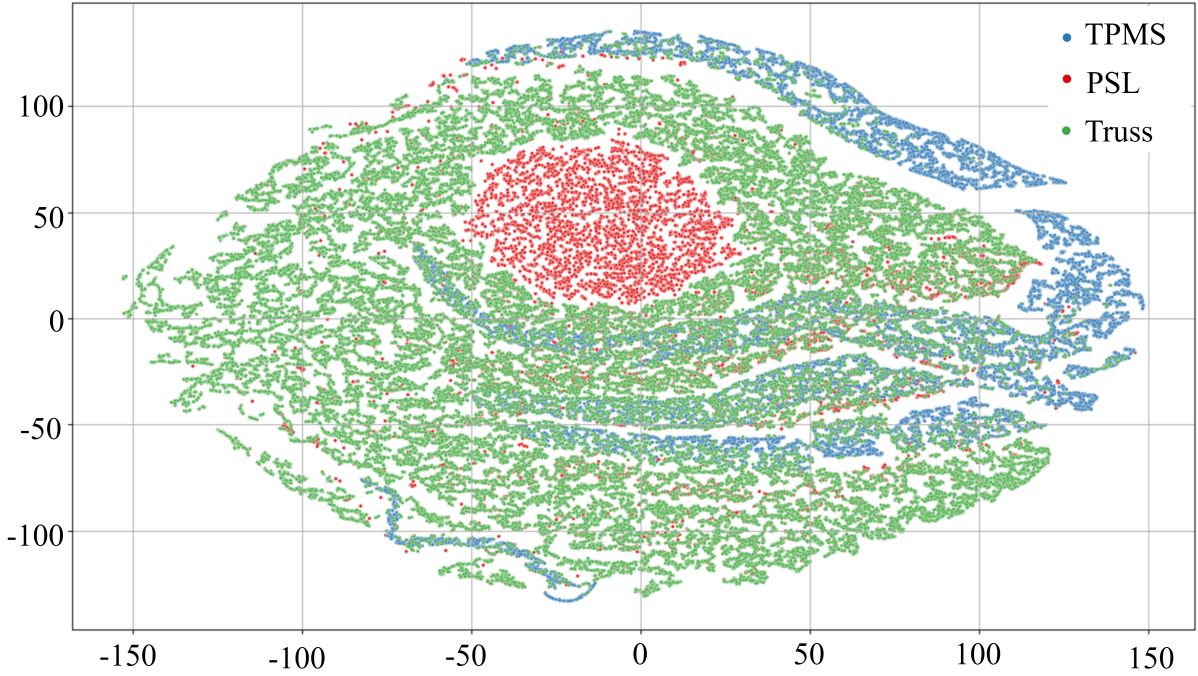}
    \caption{\textbf{T-SNE visualization of effective elasticity tensor distributions derived from TPMS, PSL, and Truss microstructures.} This figure shows a t-SNE projection of the $6 \times 6$ effective elasticity tensors ($C^H$) from three datasets: TPMS (blue), PSL (red), and Truss (green).  Each point represents a microstructure’s mechanical response, embedded from the 36-dimensional elasticity space into 2D. TPMS samples exhibit a uniform yet compact distribution, indicating limited but evenly sampled property variation.  PSL spans a broader region but clusters densely in a narrow subspace, with outliers scattered sparsely.  In contrast, Truss structures display the widest and most dispersed coverage, reflecting their topological complexity and highly diverse mechanical behaviors. This analysis highlights the increasing structural and property diversity from TPMS to Truss, underscoring the challenge of building generalizable predictors.}
    \label{fig:tsne-display}
\end{figure*}

We performed a t-SNE dimensionality reduction on the $6 \times 6$ effective elasticity tensor $C^H$ computed from all microstructure samples, projecting the data into a 2D space for visualization. As shown in \cref{fig:tsne-display}, each point corresponds to a sample in one of the three datasets: TPMS (blue), PSL (red), or Truss (green). As shown, the three datasets form distinguishable yet partially overlapping clusters, reflecting their differences in topological regularity, geometric complexity, and resulting mechanical behavior.

This visualization highlights the diversity of the benchmark datasets. The tight clustering of PSL indicates structural regularity and homogeneity in its elastic responses, while Truss and TPMS samples occupy broader, more complex regions in the feature space. These distinctions help explain the varying difficulty levels across datasets when predicting elastic properties.

\begin{table}
    \centering
    \caption{\textbf{Overview of microstructure datasets with topological and volume fraction variations.} This table summarizes the datasets used for training and evaluation. Each dataset represents periodic microstructures with varying topology and volume fraction (VF). For TPMS, topology refers to five distinct implicit surface equations (see Table~\ref{tab:A-TPMS}); for PSL and Truss, it refers to different skeletal graph structures derived from procedural designs or engineering templates. Each dataset includes separate splits for training, test, and validation. Poisson’s ratios $\nu$ are randomly sampled from $[0.2, 0.4]$, and for benchmarking against baseline methods, material properties are fixed at $E = 1$, $\nu = 0.3$. The volume fraction for all samples ranges from $10\%$ to $40\%$.}
    \scalebox{.83}{
        \begin{tabular}{lllcl}
            \toprule
             \textbf{Dataset}& \textbf{Method} & \textbf{Topology} & \textbf{Training} / \textbf{Test} / \textbf{Validation} & \textbf{VF range} \\
             \midrule
             TPMS &Tab.~\ref{tab:A-TPMS} &5& 34016 / 3000 / 3000 &$10\% - 40\%$ \\
             PSL &\cite{liu2022parametric} &26224 & 38877 / 3000 / 3000 &$10\% - 40\%$\\
             Truss & \cite{saeb2016aspects,panetta2015elastic}&26224 & 78604 / 3000 / 3000 &$10\% - 40\%$\\
           \bottomrule
        \end{tabular}
        }
    \label{tab:dataset}
\end{table}
\begin{figure}
    \centering
    \includegraphics[width=\linewidth]{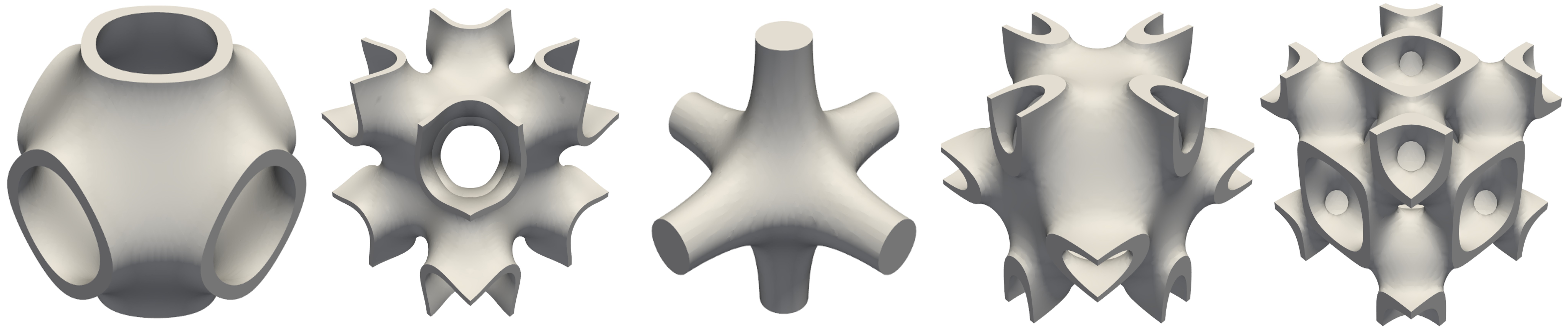}
    \leftline{ \hspace{0.4\linewidth}(a) TPMS}
    \includegraphics[width=\linewidth]{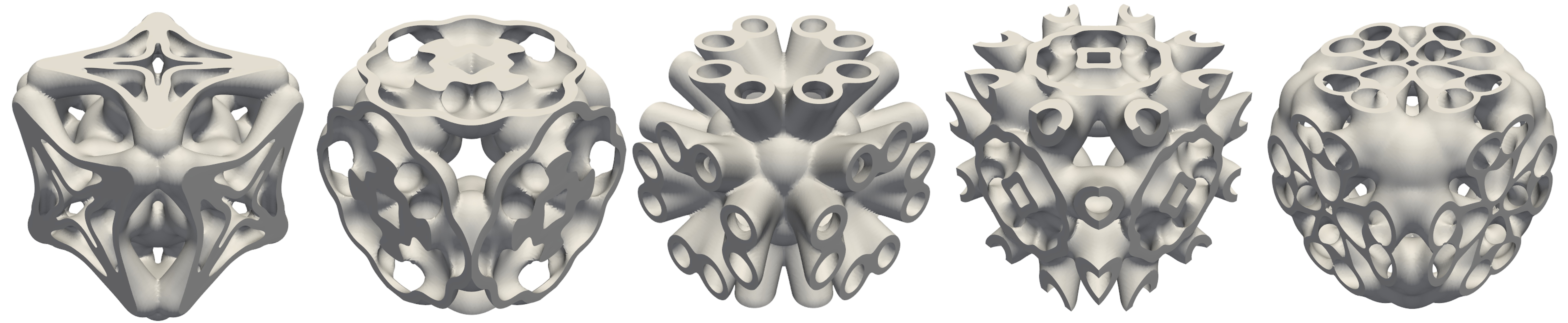}
    \leftline{ \hspace{0.4\linewidth}(b) PSL}
    \includegraphics[width=\linewidth]{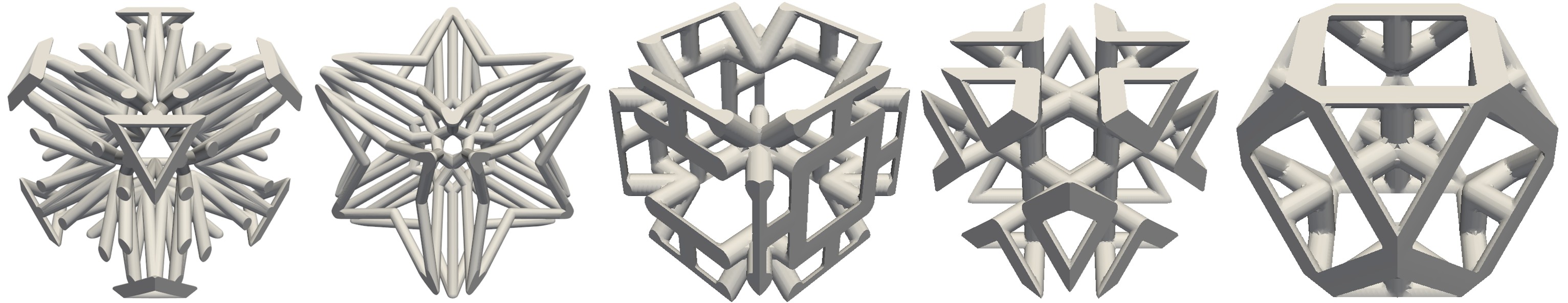}
    \leftline{ \hspace{0.4\linewidth}(c) Truss}
    \caption{\textbf{Representative microstructure samples from three datasets}}
    \label{fig:dataset-display}
\end{figure}

\section{Additional validations}
\begin{table*}
    \centering
    \caption{\textbf{Comparative analysis of prediction errors between \name and baseline methods (3D-CNN, Label-Free, and PH-Net)}. All methods operate at $n=64$ resolution. Material properties are fixed at $E=1$ and $\nu=0.3$ for fair comparison. Volume fraction (VF) ranges from $10\%$ to $40\%$, divided into three intervals: $10\%-20\%$, $20\%-30\%$, and $30\%-40\%$, with each subset containing 1,000 microstructures. For each method, maximum error ($\delta_{Max}$), minimum error ($\delta_{Min}$), and mean error ($\delta_{Mean}$) are reported.}
    \scalebox{1}{
    \begin{tabular}{ccrrr|rrr|rrr|rrr}
        \toprule
        \multirow{2}{*} {\textbf{Dataset} } & \multirow{2}{*}{\textbf{VF}} & \multicolumn{3}{c}{\textbf{3D-CNN}~\cite{rao2020three}} &\multicolumn{3}{c}{\textbf{Label-Free}~\cite{zhu2024learning}} &\multicolumn{3}{c}{\textbf{PH-Net}~\cite{peng2022ph}} &\multicolumn{3}{c}{\textbf{\name}} \\
          & & $\bm{\delta_{Max}}$ &$\bm{\delta_{Min}}$ &$\bm{\delta_{Mean}}$   & $\bm{\delta_{Max}}$ &$\bm{\delta_{Min}}$ &$\bm{\delta_{Mean}}$   & $\bm{\delta_{Max}}$ &$\bm{\delta_{Min}}$ &$\bm{\delta_{Mean}}$   & $\bm{\delta_{Max}}$ &$\bm{\delta_{Min}}$ &$\bm{\delta_{Mean}}$  \\
        \midrule
        \multirow{3}{*}{ TPMS } 
        & $10\%-20\%$ &$17.57\%$ &$6.08\%$ &$10.22\%$ &$44.81\%$ &$16.11\%$ &$22.68\% $ &$2.34\%$ &$0.88\%$ &$1.43\%$&\textbf{$\bm{0.18\%}$}&\textbf{$\bm{0.00\%}$} &$\bm{0.06\%}$\\
        & $20\%-30\%$ &$11.98\%$ &$5.22\%$ &$9.04\%$ &$32.49\%$ &$14.31\%$ &$32.49\%$ & $2.62\%$&$0.86\%$ &$1.49\%$ &\textbf{$\bm{0.09\%}$} &\textbf{$\bm{0.02\%}$} &\textbf{$\bm{0.04\%}$} \\
        & $30\%-40\%$ &$11.51\%$ & $2.10\%$ & $6.36\%$ & $19.80\%$ &$10.86\% $&$15.36\%$ &$2.53\%$ &$0.85\%$ &$1.44\%$  &\textbf{$\bm{0.12\%}$} &\textbf{$\bm{0.02\%}$} &\textbf{$\bm{0.04\%}$} \\

        \multirow{3}{*}{  PSL } 
        & $10\%-20\%$ &$106.64\%$ &$23.76\%$ &$51.87\%$ &$226.26\%$ &$46.86\%$ & $100.81\% $&$473.37\%$ &$4.92\% $&$30.08\%$ &\textbf{$ \bm{0.64\%}$}&\textbf{$\bm{0.22\%}$} &\textbf{$\bm{0.32\%}$} \\
        & $20\%-30\%$ &$111.99\%$ &$12.20\%$ &$41.47\%$ &$239.98\%$ &$43.24\%$ &$92.86\%$&$308.04\%$ &$5.52\%$ &$23.60\%$ &\textbf{ $\bm{0.55\%}$}&\textbf{$ \bm{0.19\%}$}& \textbf{$\bm{0.32\%}$}\\
        & $30\%-40\%$ &$73.73\%$ &$8.57\%$ &$27.13\% $ &$155.86\%$ &$30.54\%$ & $68.85\%$&$188.32\%$ &$4.12\%$ &$15.90\%$ &\textbf{$\bm{0.38\%}$} &\textbf{$\bm{0.14\%}$} &\textbf{$\bm{0.29\%}$} \\

        \multirow{3}{*}{Truss} 
        & $10\%-20\%$ &$197.54\%$ &$16.02\%$ &$66.94\%$ &$836.25\%$ &$54.58\%$ &$278.99\%$ &$104.04\%$ &$4.90\%$ &$65.67\%$ & \textbf{$\bm{0.84\%}$}& \textbf{$\bm{0.32\%}$} & \textbf{$\bm{0.64\%}$} \\
        & $20\%-30\%$ &$142.22\%$ &$9.86\%$ &$43.94\%$ &$463.34\%$ &$40.61\%$ & $158.90\%$ & $357.14\%$&$3.25\%$ &$36.31\%$ &\textbf{$\bm{0.73\%}$ }&\textbf{$\bm{0.19\%}$ }&\textbf{$\bm{0.51\%}$} \\
        & $30\%-40\%$ &$98.90\%$ &$7.66\%$ &$30.52\%$ &$249.64\%$ &$33.17\%$ &$94.53\%$ & $329.14\%$&$2.58\% $&$18.43\%$ &\textbf{$\bm{0.43\%}$} &\textbf{$\bm{0.10\%}$}&\textbf{$ \bm{0.38\%}$}\\
       \bottomrule
    \end{tabular}
    \label{tab:a-time}
    }
   
\end{table*}
\subsection{Prediction accuracy}
To provide a more detailed comparison of prediction accuracy, we conducted a comprehensive statistical analysis across varying volume fractions within the test sets.
Quantitative comparisons of maximum, minimum, and mean errors for each baseline method and \name are systematically summarized in \cref{tab:a-time}.

To accommodate the resolution limits of all baseline methods, the resolution was fixed at $n=64$.
For consistency, the material properties were set to $E=1$ and $\nu=0.3$ across all test cases.

\begin{figure*}
    \centering
    \includegraphics[width=\linewidth]{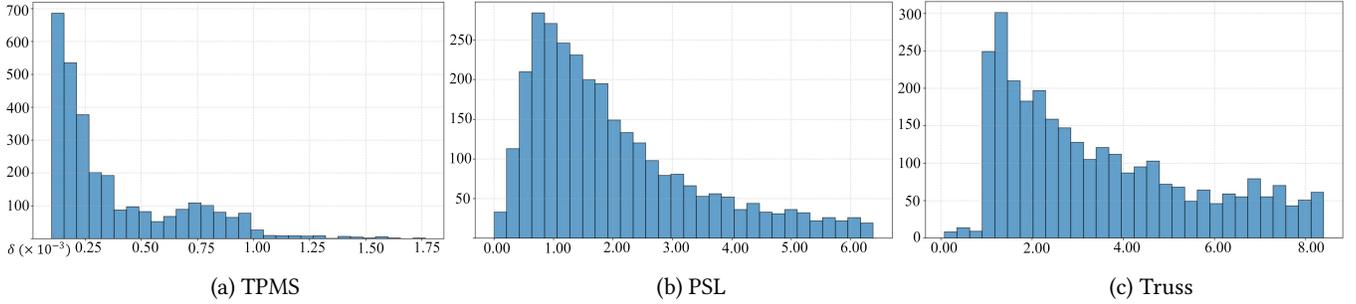}
    \leftline{ 
    \hspace{0.15\linewidth}
        (a) TPMS\hspace{0.26\linewidth}
        (b) PSL\hspace{0.28\linewidth}
        (c) Truss}
     \caption{ \textbf{Histogram of prediction error distributions for \name across (a) TPMS, (b) PSL, and (c) Truss datasets.} Each plot highlights the robustness of \name predictions across varying structural complexities.}
    \label{fig:error-display}
\end{figure*}

The error distributions of \name across the TPMS, PSL, and Truss datasets were further visualized in \cref{fig:error-display}, highlighting the robustness and stability of our predictions across diverse structural complexities.

\subsection{Runtime breakdown of \name components}
\label{A-ModuleTime}
\begin{table}[h]
    \centering
    \caption{\textbf{Runtime breakdown (s) of individual components in \name across different voxel resolutions.} Measured on a single A40 GPU. The embedded PCG solver maintains low overhead across all scales, while the sparse convolutional encoder and decoder dominate total cost at high resolutions.}
    \scalebox{0.95}{
    \begin{tabular}{l|rrrrr}
        \toprule
        \textbf{Resolution} & \textbf{Init.}& \textbf{Encoder} & \textbf{U-Net} & \textbf{Decoder} & \textbf{PCG}  \\
        \midrule
        $64$  & 0.06& 0.02  & 0.05  & 0.02  & 0.03   \\
        $128$ & 0.28& 0.10  & 0.40  & 0.11  & 0.24    \\
        $256$ & 0.02& 0.34 & 0.78  &  0.33 & 1.86    \\
        $512$ & 0.31& 0.78 & 1.95  &  0.91&  6.28   \\
        \bottomrule
    \end{tabular}
    }
    \label{tab:A-TimeBreakdown}
\end{table}
To better understand the runtime behavior of \name, we provide a detailed breakdown of the computational cost associated with each major component in the network. Specifically, we measure the average forward-pass time for the following modules: Encoder, U-Net, Decoder and PCG.

All timings are recorded on a single NVIDIA A40 GPU using a batch size of 1, averaged over 1000 test runs. The results are shown in \cref{tab:A-TimeBreakdown}.

\subsection{Comparison with larger PH-Net}
While \name achieves superior prediction accuracy, it must be noted that it also has a higher number of trainable parameters compared to baseline networks.
To rigorously assess whether the performance gains stem merely from increased network capacity, we conducted controlled experiments by scaling up PH-Net.

Specifically, we enlarged the convolutional kernel sizes and feature dimensions of PH-Net, resulting in an augmented model with approximately $14$ million trainable parameters -- comparable to \name.
The performance of this enlarged PH-Net was evaluated across the TPMS, PSL, and Truss test datasets.

As shown in \cref{tab:PHNET-BIG}, although the scaled PH-Net exhibits improved prediction accuracy relative to its original configuration, it still fails to match the precision, residual reduction, or stability achieved by \name.
The relative errors in predicting $C^H$ remain substantially high, particularly on structurally complex datasets. 
These results further confirm that the superior performance of \name stems from its integrated architectural innovations and numerical coupling, rather than merely increased model capacity.

\begin{table}
    \centering
     \caption{\textbf{Prediction performance of PH-Net scaled to approximately $14$ million trainable parameters.} Despite increased capacity, the enlarged PH-Net exhibits substantially higher prediction errors and residuals compared to \name.}
    \begin{tabular}{lrrrr}
        \toprule
        \textbf{Dataset }  & $\bm{\delta_{Max} }$ &$\bm{\delta_{Mean}}$ & $\bm{\Vert r\Vert_{Max}}$ & $\bm{\Vert r\Vert_{Mean}}$  \\
        \midrule
        TPMS &$57.46\%$  & $1.12\%$ & 24.17  & 9.81 \\
        PSL  & $28.85\%$ & $9.25\%$ & 37.83  & 15.16\\
        Truss  & $977.29 \%$ & $34.76\%$ & 36.46  & 14.88 \\
       \bottomrule
    \end{tabular}
    \label{tab:PHNET-BIG}
 \end{table}

\subsection{Inference speed of neural homogenization baselines}
\begin{table}
        \caption{\textbf{ Inference time of neural solvers on resolution $64$ Truss structures.} Although baseline methods exhibit low latency, they fail to provide physically reliable outputs (see main results). \name offers a balanced trade-off between speed and solution accuracy.}
    \scalebox{0.98}{
    \begin{tabular}{lrlr}
        \toprule
        \textbf{Network} & \textbf{Time (ms)} & \textbf{Prediction type} &$\bm{\delta_{Mean}}$ \\
         \midrule
         3D-CNN & 42.19& Predicts $C^H$ directly& $\bm{31.94\%}$\\
          Label-Free &104.81 & From $u$ to $C^H$& $\bm{288.49\%}$\\
          PH-Net & 120.77& From $u$ to $C^H$& $\bm{30.10\%}$\\
          \name & 144.07 & From $u$ to $C^H$& $\bm{0.29\%}$\\
       \bottomrule
    \end{tabular}
    }
    \label{tab:related-work-time}
\end{table}

While learning-based approaches are often claimed to be faster than traditional numerical solvers, their predictive reliability varies significantly across methods. To investigate this trade-off, we report the inference (forward pass only) of each neural method on Truss structures with $n=64$ resolution, using a single NVIDIA RTX A40 GPU.

As shown in \cref{tab:related-work-time}, although 3D-CNN, Label-Free, and PH-Net offer fast inference times, their predictions often suffer from high residuals and elastic tensor errors (see Sec.~2.2), making them unsuitable as stable solvers. In contrast, \name achieves near-real-time speed with substantially higher accuracy and robustness, validating its practical applicability in high-throughput settings.

\revision{
\subsection{Comparison of learning from neural–numerical hybrid approaches}
\begin{table}
        \caption{\revision{\textbf{CComparison of learning from neural–numerical hybrid approaches.} \textbf{Type} denotes how the neural network is combined with numerical solvers. \textbf{Max Res.} indicates the maximum supported resolution of the method on the lattice homogenization problem. All evaluations are conducted on the Truss dataset with a resolution of 64. \textbf{Pre-Time} denotes the preparation time for each test sample. Since GNP and Solver-in-the-Loop lack generalization across different structures, they require retraining for each sample, and their training time is counted into Pre-Time. \textbf{For-Time (s)} is the forward time when testing. We regard \textbf{Pre-Time} $+$ \textbf{For-Time} as the actual time cost for solving one model. $\delta_{Mean}$ and $\Vert r \Vert_{Mean}$ represent the mean error and residual over successfully solved samples in test set. Success rate is defined as the proportion of test samples solved with error below $1\%$.}}
    \begin{tabular}{lll}
        \toprule
        \textbf{Network} & GNP\cite{chengraph} &Solver-in-the-Loop\cite{um2020solver}\\
        \midrule
        \textbf{Type} & Preconditioner & Initial Guess \\
        \textbf{Max Res.} & 128 &64 \\
        \textbf{Pre-Time (s)}& 993.00 & 29600.00 \\
        \textbf{For-Time (s)}&12.43& 1.20 \\
        $\bm{\delta_{Mean}}$ &$0.02\%$ & $0.12\%$\\
        $\bm{\Vert r\Vert_{Mean}}$ & $4.14\times10^{-4}$& $3.42\times10^{-3}$ \\
       \bottomrule
    \end{tabular}
    
    \label{tab:A-LearningAlgebra}
\end{table}
Neural–numerical hybrids for linear solves largely follow three templates: learn a preconditioner, learn a component of an algebraic multigrid and learn an initial guess that is then refined by an iterative solver. We evaluate one representative method from each category on our 3D periodic lattice homogenization task. Unless noted otherwise, we use the authors’ official implementations and retain each pipeline unchanged except for the learned component they target. All experiments are run on the Truss dataset at $64^3$ resolution. The results are shown in Table~\ref{tab:A-LearningAlgebra}.
\paragraph{GNP}
We compare against the Graph Neural Preconditioner (GNP) \cite{chengraph}, which learns a preconditioner $M \approx A^{-1}$ from the coefficient matrix $A$ via a graph neural network (GNN) and uses it within flexible GMRES (FGMRES). This “learn-the-inverse” design explicitly approximates the action of $A^{-1}$ for the purpose of preconditioning. In our setting, this raises three practical issues:\\
\textbf{Resolution and memory}: even when $A$ is sparse, $A^{-1}$ is structurally dense, which is a classical fact in sparse linear algebra and underlies the need for approximate inverses. The official implementation was obtained from the public repository\footnote{\url{https://github.com/jiechenjiechen/GNP}}. As a result, the memory and compute footprints of GNP’s learned inverse-operator grow steeply; in our 3D problems we were unable to push beyond a resolution of $128^3$ without running into prohibitive GPU memory/time. \\
\textbf{Limited cross-matrix generalization}: GNP is trained for a single $A$; the authors themselves note it is unrealistic to expect one network to work for all matrices, suggesting per-matrix fine-tuning if broader coverage is desired. Consequently, when $A$ changes the preconditioner must be relearned, which introduces substantial extra time cost in our pipeline (see Table~\ref{tab:A-LearningAlgebra}, row 1). \\
\textbf{Effectiveness on our targets}: on complex-structure linear systems, GNP did not surpass our \name baseline in solution accuracy. By contrast, our method is designed to train one network that predicts the required structural properties across the full parameter range without any per-instance retraining, making the GNP strategy misaligned with our goals and constraints.
\paragraph{Solver-in-the-Loop}
We adapt "solver-in-the-loop"~\cite{um2020solver} to 3D periodic lattice homogenization. The official implementation was obtained from the public repository\footnote{\url{https://github.com/tum-pbs/CG-Solver-in-the-Loop}}. 
Unlike the original Poisson setting, which involves a divergence scalar field, our network takes as input the node-wise assembled stiffness and external load. It then predicts an initial displacement field $u_0$, which is further refined by a standard CG solver.
Training unrolls CG and backpropagates through the iterations; at evaluation we stop CG once the accuracy matches \name's result. \\
\textbf{Resolution and memory.} Dense, fully padded grids together with reverse-mode differentiation through tens of CG steps create long gradient chains, capping the feasible resolution at $64^3$ in our setup. Per-geometry training requires 5.5 h, while a trained model’s forward (network + test-time CG) is 1.2 s. \\
\textbf{Effectiveness on our targets.} The "solver-in-the-loop" does not generalize across geometries: models trained on one lattice must be retrained for another, which is incompatible with design-space sweeps. Moreover, reaching residual error with \name typically still takes $\approx40$ CG iterations. As shown in Table~\ref{tab:A-LearningAlgebra}, row 3, we obtain $0.12\%$ error and a final residual of $3.42\times 10^{-3}$. Overall, although philosophically similar (learned initializer + CG), the dense, unrolled training loop and lack of cross-geometry generalization make this solver-in-the-loop variant ill-suited for high-resolution lattice homogenization and for workflows that require rapid evaluation across many microstructures.}

\revision{Our problem imposes strict requirements—high 3D resolution, geometric periodicity, and hard physical constraints. Methods that learn an inverse effectively realize near-dense operators, which makes memory and runtime prohibitive at the resolutions we target. Approaches that learn AMG components have likewise not proven robust in our setting, exhibiting low success rates and lacking convergence guarantees. In contrast, learning an initial guess is, at present, the only pathway that satisfies these constraints. When the initializer is produced by a network that operates directly on the 3D periodic geometry and is trained with explicit physics and periodicity constraints, the subsequent CG refinement achieves the best trade-off: high accuracy and stable convergence. Accordingly, we adopt this physics- and geometry-aware initializer strategy in \name.}

\revision{
\subsection{Comparison of different numerical solvers}
    

\begin{table}[t]
    \centering
    \caption{\revision{\textbf{Performance comparison of different iterative solvers on the Truss dataset at resolution 64.} We report the performance of several classical iterative solvers under the same experimental setup. Each solver was run for 10 iterations, and we recorded three metrics: the runtime (Times), the relative error of the computed solution with respect to the reference solution ($\delta_{\mathrm{Mean}}$), and the residual norm of the linear system ($\lVert r\rVert_{\mathrm{Mean}}$). This comparison highlights the efficiency and accuracy differences among Jacobi, Gauss–Seidel (G-S), Successive Over-Relaxation (SOR), Conjugate Gradient (CG), and Preconditioned Conjugate Gradient (PCG) with Jacobi preconditioning.}}
        \begin{tabular}{l|rrr}
            \toprule
             \textbf{Solvers} & \textbf{Time (s)} & \multicolumn{1}{c}{$\bm{\delta_{Mean}}$} & \multicolumn{1}{c}{$\bm{\Vert r\Vert_{Mean}}$ } \\
             \midrule
             Jacobi &0.11 & Divergent &Divergent\\
             G-S  &873.55 & $122.31\%$&$3.03\times10^{-2}$ \\
             SOR ($\omega = 1.2$)  &955.23 & $90.55\%$&$1.04\times10^{-2}$ \\
             CG   &0.13 & $10.08\%$&$8.33\times10^{-3}$\\
             PCG (Jacobi)  &0.15 & $0.51\%$ &$3.34\times10^{-3}$\\
            \bottomrule
        \end{tabular}
    \label{tab:A-numerical-solvers}
 \end{table}
We conducted additional experiments on the Truss dataset to compare several classical iterative solvers under the same test conditions, as summarized in Table~\ref{tab:A-numerical-solvers}. The Jacobi method fails to converge because the system matrix is not strictly diagonally dominant. The Gauss–Seidel (G-S) and Successive Over-Relaxation (SOR, $\omega = 1.2$) methods are theoretically convergent, but they cannot be efficiently parallelized and exhibit very high computational costs (over 800 seconds) with relatively large errors. The Conjugate Gradient (CG) method achieves convergence with a much lower runtime (around 0.13 seconds) and acceptable accuracy ($\approx 10\%$ error), but its precision is still limited. In contrast, the Preconditioned Conjugate Gradient (PCG) method with Jacobi preconditioning maintains a similar level of efficiency while further reducing the average error to below $1\%$ and achieving the smallest residual norm.
Overall, PCG offers the best balance among convergence, computational efficiency, and numerical accuracy, and is therefore adopted in this work.
}
\revision{
\subsection{Comparison of different preconditioners}
}
\revision{
\begin{table*}[h]
\centering
\caption{\revision{\textbf{PCG preconditioners: per-sample runtime, accuracy, and residual.}
We evaluate several commonly used preconditioners—Jacobi (diagonal scaling), incomplete Cholesky with zero and level-2 fill-in (IC(0), IC(2)), and algebraic multigrid (AMG). 
We report per-sample time for training and inference under identical settings
(\emph{training} = one forward+backward pass per sample; \emph{inference} = one forward pass).
We also report the relative error to the reference solution ($\delta_{\text{mean}}$) and the mean residual norm ($\|\mathbf{r}\|_{\text{mean}}$).}}
\begin{tabular}{l|cc|cc}
\toprule
\multirow{2}{*}{\textbf{Preconditioner}} & \multicolumn{2}{c}{\textbf{Time (per sample)}} & \multicolumn{2}{c}{\textbf{Metrics}}\\
 & \textbf{Forward+Backward (s)} & \textbf{Forward (s)} &$\delta_{Mean}$ & $\Vert r\Vert_{Mean}$ \\
     \midrule
    Jacobi (\name) &0.45 & 0.12 & $0.51\%$ &$3.38\times10^{-3}$\\
    IC(0)  &4.10 & 3.18 & $0.26\%$ &$1.32\times10^{-3}$\\
    IC(2)  &7.33 & 6.29 & $0.22\%$ &$1.29\times10^{-3}$\\
    AMG    &2.41 & 1.71 & $0.38\%$ & $2.05\times10^{-3}$\\
    \bottomrule
    \end{tabular}
    \label{tab:A-preconditioner}
\end{table*}

We ablate the choice of preconditioner on the Truss dataset at resolution 64 with a fixed budget of 10 solver iterations.  
We compare Jacobi (diagonal scaling), IC(k) (incomplete Cholesky with level-of-fill $k \in \{ 0, 2\}$, implemented with cuSPARSE\footnote{\url{https://developer.nvidia.cn/cusparse}}), and AMG (algebraic multigrid, NVIDIA AmgX\footnote{\url{https://github.com/NVIDIA/AMGX}}).  
As shown in Table~\ref{tab:A-preconditioner}, the stronger preconditioners provide only modest improvements in residual and error under this regime, while their runtime is more than an order of magnitude higher than Jacobi. 
}

\revision{
The increased costs of IC and AMG arise from different sources.  For IC, the overhead stems not only from the setup phase—where symbolic analysis and numerical factorization are required—but also from the application phase, since each iteration performs two sparse triangular solves.  These solves expose limited parallelism on GPUs, are memory- and latency-bound, and become increasingly expensive as the fill level grows from IC(0) to IC(2), leading to higher runtime despite stronger conditioning.  In contrast, the main expense of AMG lies in constructing the multilevel hierarchy, including aggregation, interpolation operators, and smoothers.  Given only 10 solver iterations, this setup cost is barely amortized, which renders AMG inefficient in our regime.  Meanwhile, Jacobi has negligible setup, amounts to a single bandwidth-friendly scaling per iteration, and maps efficiently to GPUs.  Combined with the learned initializer, $u_0$, which places the solver close to the solution, Jacobi attains low residuals within 10 steps, making heavier preconditioners unattractive in this setting.
}
\revision{
\subsection{Ablation study on loss functions}
\begin{table}[t]
    \centering
    \caption{\revision{\textbf{Ablation study on different loss functions.} 
    We report the dataset preparation time (P-Time), training time (T-Time), mean error on the test set ($\delta_{\mathrm{Mean}}$), and mean residual norm ($\lVert r \rVert_{\mathrm{Mean}}$), based on the Truss dataset at resolution 64.}}
        \begin{tabular}{l|rr|rr}
            \toprule
             \textbf{Loss} & \textbf{P-Time (h)} & \textbf{T-Time (h)}& $\bm{\delta_{Mean}}$ & $\bm{\Vert r\Vert_{Mean}}$   \\
             \midrule
            Label & 44.20& 123.30 & $118.10\%$ &$3.66\times 10^{-2}$ \\
            Residual & / & 130.50 & $8.10\%$ &$4.21\times 10^{-3}$ \\
            Energy  &  / & 125.20   & $0.84\%$ &$3.46\times 10^{-3}$\\
            \bottomrule
        \end{tabular}
    \label{tab:A-loss}
 \end{table}
To evaluate the impact of different loss formulations, we conducted ablation experiments on the Truss dataset under identical training settings, with only the choice of loss function varied. Let $u$ denote the network output.}

\revision{First, we considered a label-based supervised approach with the loss function $L_{label}=MSE(u,u_{gt})$, where $u_{gt}$ is the ground-truth displacement field. This approach requires prior labeling of the dataset, which incurred significant preprocessing costs ($44.2$ hours for data preparation). Despite this effort, the final performance on the test set was unsatisfactory, with an error of $118.10\%$. These results suggest that purely label-driven training is insufficient for this task, as it fails to incorporate the necessary physical constraints of the problem.}

\revision{To address this limitation, we investigated physics-informed loss functions. Specifically, we tested a residual-based formulation, $L_{residual}=\Vert Ku-f||^2$ which enforces the equilibrium equations, and an energy-based formulation (Equation 12 in the main text), which captures the underlying variational structure of the problem. As summarized in Table~\ref{tab:A-loss}, both physics-based approaches significantly improved performance compared to the label-based method. Notably, the energy-based loss led to the most stable and accurate results, achieving the lowest mean relative error ($\delta_{Mean}=0.84\%$ and the smallest residual norm $\Vert r\Vert_{Mean}=3.46\times 10^{-3}$). These findings confirm that embedding physical knowledge into the training process is critical, and the energy-based loss provides a more robust and effective framework than either label supervision or residual minimization. 
}

\revision{
\subsection{Sensitivity to noisy data}
\begin{table}[t]
    \centering
    \caption{\revision{\textbf{Sensitivity to voxel noise.} Results on the TPMS and Truss test sets at resolution 64 under random occupancy flips with noise ratio $p$; the pre-trained model is frozen and only test voxels are perturbed. We report the mean relative error of the homogenized elasticity tensor $\delta_{Mean}$, and the solver residual norm $\Vert r\Vert$.}}
        \begin{tabular}{lr|rr}
            \toprule
             \textbf{Dataset} & \multicolumn{1}{c}{$\bm{p}$ }  & $\bm{\delta_{Mean}}$ & \multicolumn{1}{c}{$\bm{\Vert r\Vert_{Mean}}$ }  \\
             \midrule
             \multirow{5}{*}{TPMS} 
            &$0.50\%$ &$0.97\%$ & $1.03\times10^{-3}$ \\
            &$1.00\%$ &$3.11\%$ &$1.19\times10^{-3}$\\
            &$5.00\%$ &$8.43\%$ &$2.45\times10^{-3}$\\
            &$10.00\%$&$14.32\%$ &$1.47\times10^{-3}$  \\
            &$20.00\%$&$44.32\%$ &$3.60\times10^{-3}$  \\
             \midrule
             \multirow{5}{*}{Truss} 
            &$0.50\%$ & $1.37\%$ & $1.76\times10^{-3}$ \\
            &$1.00\%$ & $19.77\%$& $3.96\times10^{-3}$ \\
            &$5.00\%$ & $55.34\%$& $3.29\times10^{-3}$\\
            &$10.00\%$& $91.22\%$& $4.47\times10^{-3}$\\
            &$20.00\%$& $248.49\%$& $3.78\times10^{-3}$ \\
            \bottomrule
        \end{tabular}
    
    \label{tab:A-noise}
 \end{table}
To assess robustness, we corrupt the test voxels by independently flipping a fraction $p$ of occupied voxels ($1\to 0$, $0\to 1$), use the pretrained model, and take the clean shapes as ground truth. We report $$\delta_{Mean} = \frac{1}{n}\sum_{1}^{n} \frac{\Vert C^{H}_N(p)-C^H_0 \Vert_F}{\Vert C^H_0 \Vert_F},$$ 
averaged over $n=3000$ test instances per task. Here, $C^{H}_N(p)$ denotes the homogenized elasticity tensor of the noisy microstructure at noise ratio $p$ predicted by our network, whereas $C^H_0$ denotes the homogenized elasticity tensor of the clean microstructure obtained by numerical computation. We also report the solver residual $\Vert r\Vert_{Mean}$. Results in Table~\ref{tab:A-noise} show $\delta_{Mean}$ grows with $p$. Truss structures are markedly more sensitive to voxel noise—e.g., at $p = 1\%$ we observe $\delta_{Mean}=19.77\%$ for Truss, versus $\delta_{Mean}=3.11\%$ for TPMS. At $p=20\%$ the gap widens to $\delta_{Mean}=249.49\%$ for Truss, versus $\delta_{Mean}=44.32\%$ for TPMS. Meanwhile, $\Vert r\Vert_{Mean}$ remains around $10^{-3}$, indicating stable numerical convergence; the degradation is dominated by geometry/parameter error rather than calculation failure. We attribute the disparity to the small cross-section and connectivity fragility of thin struts—flipping even a few voxels can sever links or cause large relative thickness changes (aliasing), which amplifies downstream homogenized elasticity tensor error. Implication for practice: designs with rod-like features demand tighter manufacturing tolerances and/or preprocessing, such as simple morphological denoising or a minimum-feature-size constraint.
}

\revision{
\subsection{Non-periodic microstructures}
\begin{figure}
    \centering
    \includegraphics[width=\linewidth]{NMI/images/non-prei.svg.pdf}
    \caption{\revision{\textbf{Representative non-periodic lattice microstructure samples.} We extracted a quarter-patch and deliberately stitched such patches to form composites with broken periodicity.}}
    \label{fig:A-non-prei}
\end{figure}
To assess performance beyond periodic lattices, we synthesized a small non-periodic test set (100 microstructures). From several base periodic Truss microstructures, we extracted a quarter patch and forcibly stitched such patches to form composites with intentionally mismatched seams, thereby breaking periodicity, as shown in Fig.~\ref{fig:A-non-prei}. The same physical settings and discretization as in the main experiments were used.}

\revision{We reused the pretrained network and masked out all Peri-mapping components during inference. No additional training or fine-tuning was performed. The model achieved a mean residual of $3.28\times10^{-3}$ and a relative error of $0.42\%$, which is similar to the accuracy observed for periodic cases.
These results indicate that the \name framework retains its accuracy on non-periodic microstructures when periodicity mappings are disabled, demonstrating straightforward extensibility beyond strictly periodic designs.
}


\revision{
\subsection{Additional convergence metrics for equations solves}
\begin{table}
    \centering
     \caption{\revision{\textbf{Convergence of linear solvers across datasets.} We report the mean relative error $\delta_{Mean}$, the absolute residual $\Vert r\Vert$, the relative residual $\Vert r\Vert / \Vert f\Vert$, and the residual reduction relative to the network initialization $\Vert r\Vert/\Vert r_0\Vert$, at resolution 64.}}
    \begin{tabular}{lcccc}
        \toprule
        \textbf{Dataset } &$\bm{\delta_{Mean}}$ & $\bm{\Vert r\Vert}$ & $\bm{\Vert r\Vert / \Vert f\Vert}$ &$\bm{\Vert r\Vert/\Vert r_0\Vert}$  \\
        \midrule
        TPMS    & $0.05\%$ & $1.16\times10^{-3}$ & $8.83\times10^{-3}$  & $9.58\times10^{-2}$ \\
        PSL     & $0.31\%$ & $2.06\times10^{-3}$ & $2.51\times10^{-2}$  & $1.04\times10^{-3}$\\
        Truss   & $0.51\%$ & $3.46\times10^{-3}$ & $1.03\times10^{-2}$  & $8.57\times10^{-4}$\\
       \bottomrule
    \end{tabular}
    \label{tab:A-relative-residual}
 \end{table}

In the main text we evaluated convergence using the absolute residual. For completeness, we report two additional scale–aware metrics that are commonly used for linear systems. Consider a system $Ku=f$ and an iterative solver initialized at the neural network prediction $u_0$. Let $u_k$ be the iterate at step $k$ and $r = f-Ku_k$, $r_0 = f-Ku_0$. The results are shown in Table~\ref{tab:A-relative-residual}.
}

\section{Applications}
\subsection{Accelerated evaluation in topology optimization}
Topology optimization (TO) requires thousands of iterative structural evaluations under changing geometry, material distribution, or loading conditions. Conventional numerical solvers, such as FEM, become computational bottlenecks, especially in high-resolution 3D design spaces.

\name offers an efficient alternative by providing rapid and differentiable predictions of homogenized mechanical properties based on voxelized geometry and base materials. Once trained, \name can be integrated into the TO loop as a fast surrogate evaluator. For example, in density-based or level-set methods, intermediate candidate structures can be evaluated in milliseconds without remeshing or system assembly.

This enables a 10–100× acceleration in TO pipelines and unlocks high-resolution structural optimization in design scenarios previously limited by simulation cost. Future work may explore end-to-end integration of \name with gradient-based TO frameworks via differentiable solvers.

\subsection{High-throughput screening of mechanical metamaterials}
In metamaterial discovery, one must screen vast structural libraries -- often generated parametrically or via generative models -- for target mechanical properties (e.g., stiffness, anisotropy, Poisson's ratio). Traditional solvers are prohibitively slow for such high-throughput tasks.

\name addresses this by predicting effective elastic tensors directly from structure and base material information, achieving sub-second inference with less than $1\%$ error. This makes it feasible to evaluate tens of thousands of structures on a single GPU.

\subsection{Worst-case analysis via microstructure displacement prediction}
\label{A-WorstCase}

In addition to predicting effective material properties, \name\ supports rapid, high-fidelity displacement field inference, enabling downstream tasks such as worst-case mechanical analysis across diverse microstructures.

\begin{figure*}
    \centering
    \includegraphics[width=\linewidth]{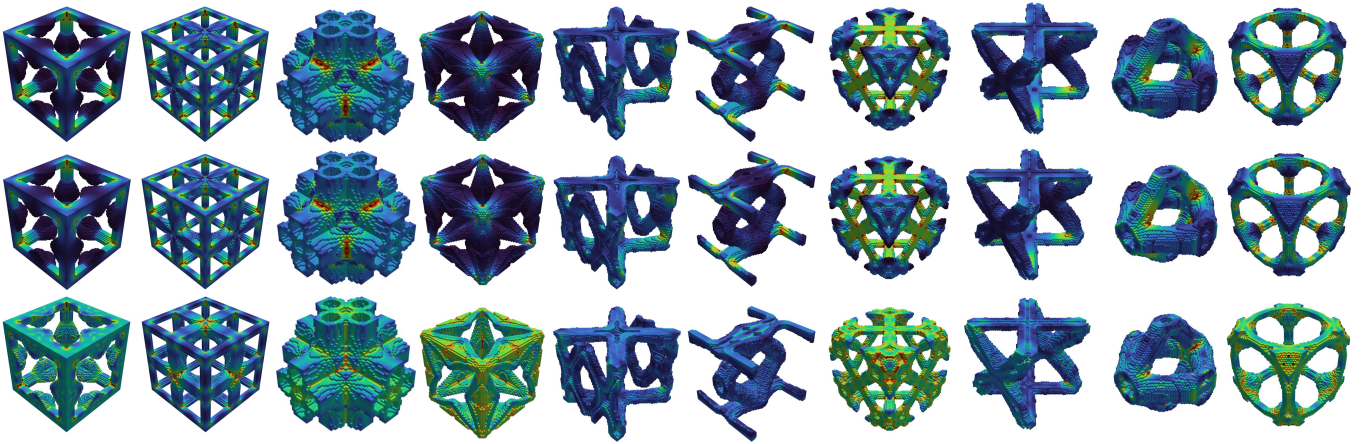}
     \caption{ \textbf{Visualization of displacement fields for worst-case analysis.} For each microstructure: Top row shows ground-truth displacement fields obtained using a validated numerical solver (referred to as~\cite{panetta2017worst}); middle row shows displacement fields predicted by \name; bottom row shows displacement fields predicted by PH-Net. \name accurately recovers detailed deformation patterns, while PH-Net exhibits significant errors.}
    \label{fig:worst-display}
\end{figure*}

To evaluate this capability, we compare the predicted displacement fields of \name\ against ground-truth numerical solutions across a representative set of microstructures with varying complexity. As shown in \cref{fig:worst-display}, these include both smooth and highly discontinuous geometries.

We adopt PH-Net~\cite{peng2022ph} as the baseline comparison method, as it is the most accurate displacement field predictor among existing neural solvers reviewed in related work. This ensures a fair and challenging benchmark for assessing our improvements.

Visual comparisons highlight that \name\ successfully reconstructs detailed displacement distributions, including subtle local deformations and discontinuities, in close agreement with the numerical ground truth. In contrast, PH-Net predictions exhibit clear deviations: displacement magnitudes are distorted, fine structures are oversmoothed, and critical stress regions are inaccurately captured—especially in geometrically intricate areas.

These results confirm that \name\ not only delivers state-of-the-art performance in homogenized property estimation, but also offers a practical and accurate tool for worst-case structural evaluation directly from input geometry—achieving substantial speedups over traditional solvers without compromising physical fidelity.



\section{Implementation details}
\subsection{Details of \name}
\label{A:Details-CGINS}
\paragraph{Network.}
\revision{\textbf{Inputs.} For each voxel v, we form two feature vectors: a material feature by unfolding its base elastic tensor $C^b\in\mathbb{R}^{6 \times 6}$ into a $36-D$ vector using a fixed row-major ordering; and a positional feature by encoding the node position $(x,y,z)$ with a Fourier feature mapping $\mathcal{F}(x,y,z)\in \mathbb{R}^{32}$ (sine–cosine features over multiple frequencies).
We concatenate both to obtain a $68-D$ per-voxel input $\mathcal{X} =[\mathtt{unfold}(C^b),\mathcal{F}(x,y,z)]\in \mathbb{R}^{68}$. 
}

\revision{
\textbf{Backbone.} The network operates over the $\mathcal{X}$. An encoder first maps the $68-D$ features to a $64-D$ latent representation, $\mathcal{Q}=\mathtt{Enc}(\mathcal{X})\in \mathbb{R}^{64}$. A U-Net~\cite{ronneberger2015u} serves as the prediction module, taking the encoder feature map as input and producing a $64-D$ feature map at the same resolution, $\mathcal{Z}=\mathtt{Unet}(\mathcal{Q})\in \mathbb{R}^{64}$. 
}

\revision{
\textbf{Outputs.} A decoder maps the $64-D$ features to the target displacement vector, $u=\mathtt{Dec}(\mathcal{Z})\in \mathbb{R}^{64}$. All convolutional layers employ ReLU activation functions to introduce nonlinearity, leveraging the Minkowski Engine~\cite{choy20194d} for efficient sparse convolution operations.
}

\revision{
\paragraph{Multi-Level Feature Refinement.}
To incorporate multi-scale context, we build a feature pyramid from the encoder output. Let level $0$ denote the  finest resolution and $L-1$ the coarsest. We obtain level-$l$ encoder features by average-pooling (downsampling): $\mathcal{Q}^l=\mathtt{Down}(\mathcal{Q}^{l-1})\in \mathbb{R}^{64}$ At each level we apply a U-Net:  $\mathcal{Z}^l=\mathtt{Unet}^l(\mathcal{Q}^l)\in \mathbb{R}^{64}$. Coarse predictions are upsampled to the next finer level and aggregated residually by element-wise addition: $\mathcal{Z}^l=\mathcal{Z}^l+\mathtt{Up}(\mathcal{Z}^{l+1})\in \mathbb{R}^{64}$. After cascading from coarse to fine, the aggregated finest-level feature map $u_0$ is passed to the decoder to produce the final $18-D$ displacement per node.
}

\paragraph{PCG iteration step selection.}
To balance numerical accuracy and runtime efficiency, we adopt different PCG iteration counts for different datasets based on their structural complexity and empirical convergence behavior. For the TPMS dataset, which exhibits smooth displacement fields and regular periodic geometry, 4 iterations are sufficient.
For the PSL dataset, which presents moderate topological complexity, 8 iterations are used to adequately suppress high-frequency residuals.
For the Truss dataset, known for its high complexity and discontinuities, we employ 10 iterations to ensure stable and accurate convergence.

These iteration counts are used during both training and inference. They are chosen to strike a practical balance between computational cost and solver fidelity.

\revision{
\paragraph{Architectural specifics.} Kernel sizes, channel widths at each stage, normalization and activation choices, pooling factors, and up/down-sampling operators for all modules (encoder, U-Nets at each level, and decoder) are provided in Table 1. We use standard concatenation (cat) for feature fusion and element-wise addition for residual aggregation across levels.
}

\subsection{PCG implementation details}
\label{A-PCG}
\paragraph{Algebra-based PCG (for $n=64$ and $n=128$)}

For low to moderate resolutions ($n=64$ and $n=128$), we implement an algebra-based PCG solver that operates directly on assembled sparse matrices $K$ and $f$.
This approach follows the classical conjugate gradient method with Jacobi preconditioning.

While assembling the full stiffness matrix $K$ and load vector $f$ incurs non-negligible memory and computational overhead, this strategy benefits from extremely fast per-iteration convergence, making it well-suited for smaller problem sizes.

The pseudo-code for the algebra-based PCG is provided below:

\begin{algorithm}[h] 
\caption{Algebra-Based PCG Solver for $Ku = f$} 
\begin{algorithmic}[1] 
    \REQUIRE Sparse matrix $K$, right-hand side $f$, initial guess $u_0$, maximum iterations $k$ 
    \STATE Compute \revision{Jacobi} preconditioner: $M = \text{diag}(K)$ 
    \STATE Initialize residual: $r_0 \leftarrow f - Ku_0$ 
    \STATE Apply preconditioner: $z_0 \leftarrow M^{-1}r_0$ (element-wise division) 
    \STATE Initialize search direction: $p_0 \leftarrow z_0$ 
    \FOR{$i = 0,1,\dots,k-1$} 
        \STATE $\alpha_i \leftarrow \dfrac{r_i^T z_i}{p_i^T K p_i}$ 
        \STATE $u_{i+1} \leftarrow u_i + \alpha_i p_i$ 
        \STATE $r_{i+1} \leftarrow r_i - \alpha_i K p_i$ 
        \STATE $z_{i+1} \leftarrow M^{-1}r_{i+1}$ 
        \STATE $\beta_i \leftarrow \dfrac{r_{i+1}^T z_{i+1}}{r_i^T z_i}$ 
        \STATE $p_{i+1} \leftarrow z_{i+1} + \beta_i p_i$ 
    \ENDFOR 
    \STATE \textbf{return} $u_k$ 
\end{algorithmic} 
\end{algorithm}

\paragraph{Geometry-based PCG: Memory-Efficient Local Stencil Implementation}
\label{A-G-PCG}
\begin{figure}
    \centering
    \includegraphics[width=.8\linewidth]{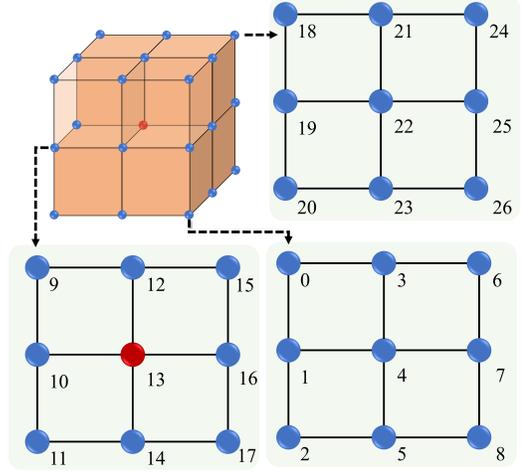}
    \caption{\textbf{Local stencil configuration for geometry-based PCG solver.} The red node denotes the target point being updated. Only its surrounding 26 nodes (in blue) and itself within a $3\times3\times3$ stencil contribute to the update, significantly reducing memory overhead. The figure decomposes the 3D neighborhood into three planar slices with labeled indices from 0 to 26, which define the local connectivity used during the implicit matrix-vector product.
}
\label{fig:G-pcg}
\end{figure}
At high spatial resolutions such as $256^3$ and $512^3$, assembling the global stiffness matrix $K$ explicitly—required for algebra-based PCG—incurs prohibitive memory costs. To overcome this, we implement a geometry-based PCG variant that circumvents matrix assembly entirely by exploiting the inherent local structure of finite element stencils.

Unlike the algebraic formulation, which relies on precomputed sparse matrices for computing matrix-vector products $Kp_i$, the geometry-based method performs this operation on the fly by iterating over each node's local 3D neighborhood. As shown in \cref{fig:G-pcg}, each node (in red) is coupled with its 26 adjacent neighbors (in blue) within a $3 \times 3 \times 3$ voxel stencil.

During each PCG iteration, instead of directly multiplying a sparse global matrix $K$ with the search direction $p_i$, we traverse each node’s local 27-node stencil, compute element-wise contributions from precomputed local stiffness blocks $K_e$, and accumulate these contributions to evaluate $Kp_i$ implicitly.

This localized computation dramatically reduces memory usage while preserving the directional update quality required for PCG convergence. Furthermore, periodic boundary conditions are enforced via modular indexing of neighbor nodes, maintaining consistency with the homogenization problem’s physical assumptions.

Importantly, the rest of the PCG steps—including residual updates, scalar coefficients ($\alpha$, $\beta$) computation, and preconditioning—remain unchanged from the algebra-based implementation described previously. The only modification lies in how $Kp_i$ is evaluated, replacing explicit sparse matrix multiplication with dynamic local stencil aggregation.

In practice, the choice between algebra-based and geometry-based PCG is adaptively determined based on resolution and memory considerations.

\subsection{Training details}
\revision{
Training is conducted in a physics-based, self-supervised manner based on the principle of minimum potential energy (see Eq. 3 in the main text), eliminating the need for ground-truth displacement supervision.
The energy-based loss further includes a penalty term on the global mean displacement to enforce boundary constraints. The weight coefficient $\lambda$ is adaptively scaled relative to the learning rate as $\lambda = 10^{-2} \times \text{lr}$, ensuring balanced optimization throughout training.}

\revision{We adopt the Adam optimizer with an initial learning rate of $3 \times 10^{-4}$, scheduled using cosine decay with a warm-up phase of 100 steps. Each model is trained for 150 epochs with a batch size of 16 (training) and 16 (evaluation), using full precision (torch.float32).}

\revision{The models are trained and evaluated on three datasets (Truss, PSL, and TPMS; see Sec.~\ref{A-Dataset} for details). Each model is trained for 150 epochs. Each epoch takes approximately 1.5 h on Truss and 0.75 h on PSL and TPMS. Training is performed on eight NVIDIA A100 GPUs with 80 GB memory, while testing is conducted on NVIDIA A40 GPUs. The implementation is based on PyTorch 1.13 and MinkowskiEngine 0.5.3, with the random seed fixed to 0 for reproducibility.
}


\subsection{Baseline specifications}
\label{sec:supp-baseline}

All baseline neural network methods were retrained on the Training sets specified in \cref{tab:dataset} and evaluated on the corresponding Test sets, consistent with the experimental setup for \name.
For numerical solver benchmarks, performance metrics were computed exclusively on the structural configurations listed in the Test datasets of \cref{tab:dataset}.

\paragraph{3D-CNN}
The 3D-CNN method was reproduced using the publicly available code repository provided by Rao et al.~\cite{rao2020three}\footnote{\url{https://github.com/Raocp/3D-ConvNeuralNet-material-property-prediction}}.
To ensure consistency across all evaluations, the input microstructure resolution was standardized to $64^3$ voxels.

\paragraph{Label-Free}
Since no open-source implementation was provided, the Label-Free method was reproduced according to the architectural details and hyperparameters specified in Zhu et al.~\cite{zhu2024learning}, based on the schematic in Fig.~7 and training settings outlined in Sec.~5.1 of their paper.
The model was trained for 200 epochs using the same training datasets.
\paragraph{PH-Net}
PH-Net was implemented using the official code repository published by Peng et al.~\cite{peng2022ph}\footnote{\url{https://github.com/xing-yuu/phnet}}.
Similar to other baselines, the input resolution was set to $64^3$ voxels to ensure fair comparisons.

\paragraph{AmgX}
The AmgX solver, developed by NVIDIA, is part of the CUDA-X GPU-accelerated library suite~\cite{AMGX}.
The official implementation was obtained from the public repository\footnote{\url{https://github.com/NVIDIA/AMGX}}.
In our experiments, we adopted the official \texttt{AMGX-LEGACY-CG} configuration for all evaluations.
\paragraph{Homo3D}
The Homo3D solver, proposed by Zhang et al.~\cite{zhang2023}, was reproduced using the publicly available code repository\footnote{\url{https://github.com/lavenklau/homo3d}}.
All Homo3D experiments were conducted in FP32 precision mode, consistent with the default settings recommended by the authors.

\section{Proofs}

\subsection{Convergence rate analysis of PCG}
\label{A-ConvergenceRate}

The effectiveness of \name relies on its ability to provide a high-quality initial guess for the preconditioned conjugate gradient (PCG) solver. In PCG, the quality of the initial guess directly impacts the convergence speed: a better initialization leads to faster residual reduction under a fixed number of iterations.

To formally characterize this relationship, we consider the linear system
\begin{equation}
K u = f,
\end{equation}
where $K \in \mathbb{R}^{3n \times 3n}$ is a symmetric positive definite stiffness matrix. To accelerate convergence, we apply PCG with a Jacobi preconditioner $M = \mathrm{diag}(K)$, leading to the preconditioned system:
\begin{equation}
\tilde{K} u = \tilde{f}, \quad \text{where} \quad \tilde{K} = M^{-1} K.
\end{equation}

Let $u^\ast$ be the exact solution and $u_0$ the initial guess. The initial error is $e = u_0 - u^\ast$, and its projection energy in the $\tilde{K}$-norm is defined as:
\begin{equation}
\Vert e \Vert_{\tilde{K}}^2 = e^\top \tilde{K} e.
\end{equation}

To understand how this energy governs convergence, we expand $e$ in the eigenbasis of $\tilde{K}$. Let $\tilde{K} = Q \Lambda Q^\top$, where $Q = [v_1, \dots, v_n]$ contains the orthonormal eigenvectors and $\Lambda = \mathrm{diag}(\tilde{\lambda}1, \dots, \tilde{\lambda}n)$ are the corresponding eigenvalues sorted in ascending order. Then:
\begin{equation}
e = \sum_{i=1}^n \alpha_i v_i, \quad \text{with} \quad \alpha_i = v_i^\top e,
\end{equation}
and the projection energy becomes:
\begin{equation}
\Vert e \Vert{\tilde{K}}^2 = \sum_{i=1}^n \tilde{\lambda}_i \alpha_i^2.
\end{equation}

This energy characterizes how much the initial error aligns with different spectral components of the preconditioned system. Since PCG reduces high-frequency (large $\tilde{\lambda}_i$) components more efficiently, convergence is dominated by the slow-to-converge low-frequency (small $\tilde{\lambda}_i$) modes. If $\alpha_i$ is large in these directions, convergence slows down.

To approximate this behavior in practice, we compute a truncated projection energy using only the smallest 20 eigenvalues of $\tilde{K}$:
\begin{equation}
\Vert e \Vert_{\tilde{K}{20}}^2 = \sum_{i=1}^{20} \tilde{\lambda}_i \alpha_i^2,
\end{equation}
which serves as a surrogate for alignment with the slowest-converging directions. A smaller value indicates that the network output is better aligned with the fast subspace of the solver, resulting in faster convergence.

In our experiments (\revision{see Fig.~3 in the main text}), \name with embedded PCG consistently produces significantly lower projection energy than both PH-Net and classical zero initialization. This confirms that joint training with the solver enables the network to learn spectral-aware solutions that are more compatible with iterative convergence behavior.

\end{document}